\documentclass[useAMS,usenatbib]{mn2e}

\usepackage{graphicx}
\usepackage{graphics,float}                
\usepackage{epsf}

\usepackage{amsmath}                                
\usepackage{amssymb}
\usepackage{picinpar}                               
\usepackage{relsize,setspace,subfigure}
\usepackage{tabularx}
\usepackage[innercaption]{sidecap}                  
\usepackage[figuresright]{rotating}
\usepackage{url}

\usepackage{natbib}      

\def\hmpc{h^{-1} {\rm Mpc}}

\def\textindent#1{\indent{#1\enspace}\ignorespaces}
\def\itemitem{\par\indent \hangindent2\parindent \textindent}

\newcommand{\cgal}{\texttt{CGAL}\ }
\newcommand {\pscz} {PSC$z$\ }
\newcommand {\nbody} {$N-$body\,}
\def\hmpc{~h^{-1} {\rm Mpc}}

\def\kms{\ {\rm km~s^{-1} }}
\def\kmsmpc{\ {\rm km~s^{-1}~Mpc^{-1} }}
\def\lessapprox{\,\raise 0.6ex\hbox{$<$}\kern -0.75em\lower 0.47ex
    \hbox{$\sim$}\,}
\def\largapprox{\,\raise 0.6ex\hbox{$>$}\kern -0.75em\lower 0.47ex
    \hbox{$\sim$}\,}

\title[DTFE analysis of the \pscz local Universe]{DTFE analysis of the \pscz local Universe:\\ 
Density Field and Cosmic Flow}

\author[Romano-D\'{\i}az \& van de Weygaert]{Emilio Romano-D\'{\i}az$^{3,2,1}$ \& 
Rien van de Weygaert$^1$\thanks{E-mail: emilio@pa.uky.edu (ERD); weygaert@astro.rug.nl (RvdW)}\\
  $^{1}$Kapteyn Astronomical Institute, University of Groningen, P.O.
  Box 800, 9700 AV, Groningen, The Netherlands.\\
  $^{2}$Racah Institute of Physics, Hebrew University, Jerusalem, 91904, Israel\\
  $^{3}$Dept. Physics \& Astronomy, Univ. Kentucky, Lexington, KY 40506-0055}

\begin{document}

\date{Accepted .... Received ...; in original form ...}

\pagerange{\pageref{firstpage}--\pageref{lastpage}} \pubyear{2007}

\maketitle

\label{firstpage}

\begin{abstract}
  We apply the Delaunay Tessellation Field Estimator (DTFE) to
  reconstruct and analyze the matter distribution and cosmic velocity
  flows in the Local Universe on the basis of the \pscz galaxy
  survey. The prime objective of this study is the production of
  optimal resolution three-dimensional maps of the volume-weighted
  velocity and density fields throughout the nearby Universe, the
  basis for a detailed study of the structure and dynamics of the
  cosmic web at each level probed by underlying galaxy sample. Fully
  volume-covering three-dimensional maps of the density and
  (volume-weighted) velocity fields in the cosmic vicinity, out to a
  distance of 150 $\hmpc$, are presented. Based on the Voronoi and
  Delaunay tessellation defined by the spatial galaxy sample, DTFE
  involves the estimate of density values on the basis of the volume
  of the related Delaunay tetrahedra and the subsequent use of the
  Delaunay tessellation as natural multidimensional (linear)
  interpolation grid for the corresponding density and velocity fields
  throughout the sample volume.  The linearized model of the spatial
  galaxy distribution and the corresponding peculiar velocities of the
  \pscz galaxy sample, produced by \cite{branchini99}, forms the input
  sample for the DTFE study. The DTFE maps reproduce the high-density
  supercluster regions in optimal detail, both their internal
  structure as well as their elongated or flattened shape. The
  corresponding velocity flows trace the bulk and shear flows marking
  the region extending from the Pisces-Perseus supercluster, via the
  Local superclusters, towards the Hydra-Centaurus and the Shapley
  concentration. The most outstanding and unique feature of the DTFE
  maps is the sharply defined radial outflow regions in and around
  underdense voids, marking the dynamical importance of voids in the
  Local Universe. The maximum expansion rate of voids defines a sharp
  cutoff in the DTFE velocity divergence pdf. We found that on the
  basis of this cutoff DTFE manages to consistently reproduce the
  value of $\Omega_m \approx 0.35$ underlying the linearized velocity
  dataset.
\end{abstract}



\section{Introduction}
The measurement and analysis of the peculiar velocities of galaxies
forms a major probe of the cosmic structure formation process. The
galaxy velocities reflect the large-scale matter flows which according
to the gravitational instability scenario for cosmic structure
formation go along with the formation and emergence of structure in
the Universe. They are the response to the gradually unfolding
underlying inhomogeneities in the cosmic matter distribution. As a
result, the comparison of the induced cosmic flows on Megaparsec
scales with the underlying matter distribution represents a
potentially powerful instrument to further our insight into the
dynamics of cosmic structure formation process and in inferring the
values of a variety of key cosmological parameters.

Over the past two decades a major effort has been directed towards
compiling large samples of galaxy peculiar velocities (see
\cite{dekel94} and \cite{strwill95} for reviews of the subject). By
opening up the window on the dynamics of the cosmic structure
formation process the analysis of these catalogues of galaxy peculiar
velocities has lead to enormous progress in our understanding of the
process. This is particularly true for scales larger than $\gtrsim
10h^{-1}\hbox{Mpc}$, scales on which structure formation still resides
in the linear phase of development. In particular the Mark III
catalog, with an effective depth $\approx 60h^{-1}\hbox{Mpc}$, stands
as a landmark achievement \citep{willick1997}.

Despite the successes progress has remained limited given the sizeable
random and systematic errors that beset catalogues of peculiar
velocities of galaxies and clusters. It allowed the mapping of the
cosmic flow field in only a rather restricted volume of the nearby
Universe.  Perhaps even more significant is that the low quality of
the peculiar velocity data, in combination with their inhomogeneous
and sparse spatial distribution, has prevented the development of an
equally compelling and complete view of cosmic dynamics at the higher
spatial resolution needed to probe and understand the dynamics of
nonlinear structures. It is on these scales of a few Megaparsec that
the structure formation process has progressed to the more advanced
quasi-linear stage and has left an imprint of genuinely recognizable
cosmological structures and patterns. The characteristic flattened and
filamentary features of the cosmic web, as well as the underdense void
regions, are particularly outstanding examples. As yet it has not been
possible to map the flows in and around these emerging structures.

Progress in the study of large scale motions will not only depend on
the availability of a substantially larger, better defined and
considerably more accurate sample of galaxy peculiar
velocities. Equally important will be the use of a more sophisticated
machinery to handle the discrete, sparse and usually inhomogeneously
sampled galaxy velocity datasets. It does remain a challenge to
transform these into properly defined velocity maps throughout the
surveyed volume. It brings to the fore the issues of interpolation of
discrete data to any location within a given volume as well as the
issue of the choice for a proper filtering prescription. As a
consequence of the specific choices and assumptions involved with a
particular method not all information inherent in the dataset is
preserved, and the resulting analysis does usually not include all
available large and small scale properties of the peculiar velocity
field. As for the first issue, the data interpolation, the measured
peculiar velocities need to be interpolated into regions devoid of
data.  A conventional approach is that of interpolation of the
discrete data set to regular grid locations by means of a fixed
interpolation kernel. A notable example is that of the TSC kernel
\citep[see][]{hockeast}. It is straightforward to show, however, that
this will result in a mass-weighted quantity~\citep{bernwey96}.  This
has confused comparisons with analytically derived results, as most of
these concern volume-weighted quantities. The second aspect is that of
the filter choice for the interpolation procedure. One may apply a
{\it fixed}, {\it global} or {\it local} filter kernel. For sensibly
defined field values it is important that data get filtered over an
appropriately large volume. Amongst other, shot-noise effects should
be suppressed. Conventionally, the galaxy peculiar velocities are
filtered on a uniform fixed scale set by the requirement that also the
sparsest sampled regions are properly averaged over. The resulting
velocity map is one in which all the velocity information on scales
smaller than the filter size has been filtered out.  The most widely
applied example is that of Gaussian filters
\citep[e.g.][]{bertschinger90, fisher1995, bernardeau95, baker98,
branchini99, dekel99}.

In this study we investigate the performance of the {\it Delaunay
  Tessellation Field Estimator} -- DTFE -- towards recovering the
spatial density field and velocity flow in the nearby Universe.  DTFE
is a self-adaptive density estimation and multidimensional
interpolation scheme which does not make use of any artificial,
user-specified filtering \cite{schaapwey2000}. Currently DTFE is the
most elaborate tessellation-based filter and interpolation scheme
available to the field of astrophysics and astronomy (see
sect.~\ref{sec:nnadapt}). It belongs to a wider generic class of
Natural Neighbour interpolations schemes. As a linear multidimensional
field interpolation scheme DTFE may be regarded as the linear
first-order equivalent of the higher order Natural Neighbour
algorithms (NN-neighbour) for spatial interpolation
\cite[see][]{sibson1981,okabe2000}.

DTFE and NN-neighbour schemes are based upon the use of the Voronoi
and Delaunay tessellations of a given spatial discrete point set
\citep[][]{voronoi1908,delaunay1934} (see \cite{okabe2000} for an
excellent and extensive overview). Voronoi and Delaunay tessellations
epitomize a purely locally defined division of space. Their
self-adaptive nature concerns both spatial resolution and local
geometry, to which they adjust through their definition as region of
influence amidst their {\it natural neighbours}. DTFE uses the high
level of sensitivity of the Voronoi/Delaunay tessellation to the local
point distribution to produce an estimate of the local density at each
sample point \citep[][]{weygaert91,weygaert94}. For telling
illustrations see \cite{schaapwey06a}.  Subsequently, it uses the
Delaunay tessellation as multidimensional spatial interpolation
grid. Unique within the context of NN-neighbour techniques. DTFE
includes the explicit extension towards the self-adaptive
determination of density fields on the basis of the spatial
distribution of the point set itself, defined by
\cite{schaapwey2000}. Also important is that its linear nature
ascertains its applicability to large data sets involving more than a
million points.

We apply DTFE to the study of the spatial mass distribution and
velocity field in the local Universe on the basis of a sample of
galaxy positions and velocities inferred from the \pscz catalog
\citep{branchini99}. A uniform, densely sampled and fully
volume-covering sample of galaxy peculiar velocities in the Local
Universe does not (yet) exist. We take the alternative of using the
galaxy velocities and positions inferred from the uniform all-sky
\pscz galaxy redshift sample. The \pscz galaxy sample is the last and
best defined galaxy redshift selection from the IRAS catalog
\citep{saunders2000}. This context guarantees the near uniform and
well-defined depth and sky coverage essential for a self-consistent
reconstruction of density and velocity fields within its volume of the
nearby Universe \citep{branchini99, schmoldt1999, branchini2001,
teodoro2003}.  Although this involves a reconstruction on the basis of
linear perturbation theory and the velocity field therefore does not
contain proper nonlinear components the spatial patterns outlined by
the galaxies' positions do provide a reasonable spatial configuration
to test DTFE over its ability to recover patterns in the velocity
field.

This case study will demonstrate the virtues and potential of the DTFE
method in the analysis of observational data, in particular in the
capacity of DTFE to reveal intricate anisotropic patterns and that of
rendering the multiscale character of both density and velocity
fields. The objective of this study is first and foremost the
production of maps of the volume-weighted velocity and density fields
throughout the nearby Universe. The self-adaptive nature of the DTFE
technique yields an optimal resolution of the most interesting
features, and this will enable the identification of characteristic
patterns in the flow field and of their principal sources in the local
density field. By default -- to the DTFE method -- this includes shear
flows, (super)cluster infall and void outflow. By illuminating their
relationship to the nearby and surrounding large scale structures the
DTFE procedure will help in increasing our understanding of the
dynamics of the formation of the various structural components of the
cosmic web \citep{bondweb96} \citep[also see][]{weyhellas2001}.

An additional aspect of the presented study is the DTFE performance in
a non-uniformly sampled galaxy distribution. Because the \pscz galaxy
catalog on which our analysis is based involves a flux-limited
selection of galaxies the sampling density gradually decreases
outward. While DTFE is capable of correcting the density values
accordingly it also introduces a differential resolution scale along
the reconstructed density and velocity field maps. While DTFE
guarantees the preservation of an optimal share of information on the
velocity field, this will automatically diminish with the outwardly
decreasing selection function and the corresponding decrease in
sampling density. The spectral coverage of the reconstructed DTFE
fields will therefore be seriously affected towards the outer region
of the \pscz volume, the velocity field reconstructions less so as
they are dominated by lower frequency components.

We introduce the \pscz sample in section 2, along with a short
description of the linearization process to produce the sample of
galaxy positions and (peculiar) velocities. This is followed by an
outline of the DTFE procedure for density and velocity fields in
section 3. A cosmographic description of the spatial structures within
the \pscz volume of the local Universe is contained in section 4,
after which in section 5 we turn to the velocity flow field within the
same volume. The issue of the differential spatial resolution of the
DTFE density and velocity maps is discussed at some length in section
6, while we describe some four specific features in the Local Universe
in section 7. Finally, we discuss the spatial distribution of the
velocity divergence field and its relation with the density
distribution. Given the rather scant information on the velocity shear
field we briefly treat its appearance in section 9. Section 10
concludes our study with a summary and a description of prospects and
relation to other work.


\section{\pscz:\\ 
\ \ \ \ \ galaxy positions and velocities} 
\label{sec:pscz}
For the study of the structure and kinematics of the local cosmic 
neighbourhood we base outselves upon the \pscz catalog. 

\subsection{The \pscz~sample}
The IRAS-\pscz ~catalog \citep{saunders2000} is an extension of the
$1.2$-Jy catalog \citep{fisher1995}. It contains $\sim 15 ~500$
galaxies with a flux at $60 \mu$m larger than $0.6$-Jy.  For a full
description of the catalog, selection criteria and the procedures used
to avoid stellar contamination and galactic cirrus, we refer the
reader to \cite{saunders2000}.  For our purposes the most important
characteristics of the catalog are the large area sampled ($\sim 84
\%$ of the sky), its depth with a median redshift of $8~500 \kms$, and
the dense sampling (the mean galaxy separation at $10~000 \kms$ is
$\langle l \rangle = 1~000 \kms$). It implies that PSCz contains most
of the gravitationally relevant mass concentrations in our local
cosmic neighbourhood, certainly sufficient to explain and study the
cosmic flows within a local volume of radius $\sim 120 \hmpc$.

Because of the flux-limited nature of the \pscz ~catalog, there is a
decrease in the objects' sampling as a function of distance from the
observer.  This is quantified by the radial selection function of the
catalog, $\psi(r)$, where the selection function is defined as the
fraction of the galaxy number density that is observed above the flux
limit at some distance $r$. To ascertain the proper number density of
objects each galaxy is weighed by the inverse of the selection
function (see eqn.~\ref{eq:densvornu}). For the selection function we
used the expression $\psi(r)$ described in \cite{branchini99}.

In addition to this selection function correction, further necessary
corrections are that for the $16 \%$ of the missing sky devoid of
data, due to the cirrus emission and unobserved areas, and that for
redshift distortions. Here we have followed \cite{branchini99} in
using their spatially reconstructed \pscz ~catalog.

\subsection{\pscz: from redshift space to physical space}
To translate the redshift space distribution of galaxies in the PSCz
catalog into galaxy positions and velocities in real space we use base
our study on a galaxy sample which has been processed by the linear
theory for gravitational instability~\citep{peebles80}.

Our sample has been obtained along the lines of {\it method I}
described in \cite{branchini99}. The method is a specific realization
of an iterative technique to minimize redshift-space distortions
\citep{yahil1991}. For a detailed exposition of the method we refer to
\cite{branchini99}. Starting with the original redshift distribution
as input and computing the corresponding peculiar velocity fields, the
spatial position of each galaxy is corrected accordingly. This is
repeated in each iterative step until the spatial distribution of the
galaxies and the inferred peculiar velocities are mutually
consistent. In order to minimize the uncertainties derived from the
lack of information on scales larger than the \pscz~catalog, the
velocity predictions in the PSCz region were made in the Local Group
frame [LG].

In order to translate the distribution of galaxies into a
gravitational field, and implied velocity field a value for the
$\beta$ parameter, the ratio between the linear velocity field factor
$f(\Omega)$ \citep{peebles80} and the (supposedly) linear bias factor
$b$ between galaxy and mass density, needs to be assumed. A value of
$\beta=0.5$ has been adopted. The reliability of the corresponding
modelled density and velocity has been confirmed in several studies,
including density-density and velocity-velocity comparisons with other
surveys \citep[e.g.][]{branchini2001,zaroubi2002}. While the
reconstruction procedure assumes linearity, one may expect its
validity to extend into the early nonlinear stages
\citep{branchini99}. At hindsight, this claim may indeed be confirmed
given the recovery by our DTFE technique of well-behaved mildly
nonlinear density and velocity divergence distribution functions (see
fig.~\ref{fig:pscz_den-div_pdf} and sect.~\ref{sec:divvpdf}).

An important condition for a consistent linear reconstruction of the
velocity and density field is that the density fluctuations should
have a small amplitude.  In order to ascertain this the gravity field
has been smoothed with a tophat filter whose radius is $500 \kms$
within the inner $50\hmpc$ and subsequently increases following the
average intergalaxy distance (see fig.~\ref{fig:pscz_intergal}).  An
additional implicit requirement for the linearization process is that
the vorticity modes in the peculiar velocity field are to be
minimized. For the well-defined and representative sample of PSCz
catalog this is accomplished for $R\approx 5\hmpc$. For comparison, it
may be worthwhile to realize that a Gaussian kernel of $12 \hmpc$ was
employed by \cite{dekel99} to obtain a succesfull and representative
linearization of the measured velocity field of the Mark III catalog
\citep{willick1997}.

A complication for the linearization procedure is that of the velocity
field around high density regions. In the immediate surroundings of
clusters of galaxies, the inflow of matter becomes so fast that they
turn into triple-valued redshift regions. \cite{emiliophd} addressed
this problem by considering two different samples. In one sample the
triple-valued regions were collapsed, in the other they were
not. Results showed that the implicit smoothing procedure of the
linearization procedure minimizes this effect. Differences between
collapsed and uncollapsed samples are less than $10 \%$ for the bulk
flow and velocity shear components, and consistent at the $1\sigma$
level. Here the velocity model which leaves the triple-valued regions
uncollapsed has been followed.

\begin{figure*}
     \begin{minipage}{\textwidth}
     \vskip -0.5truecm
    \begin{center}
      \mbox{\hskip 0.3truecm\includegraphics[width=0.9\linewidth]{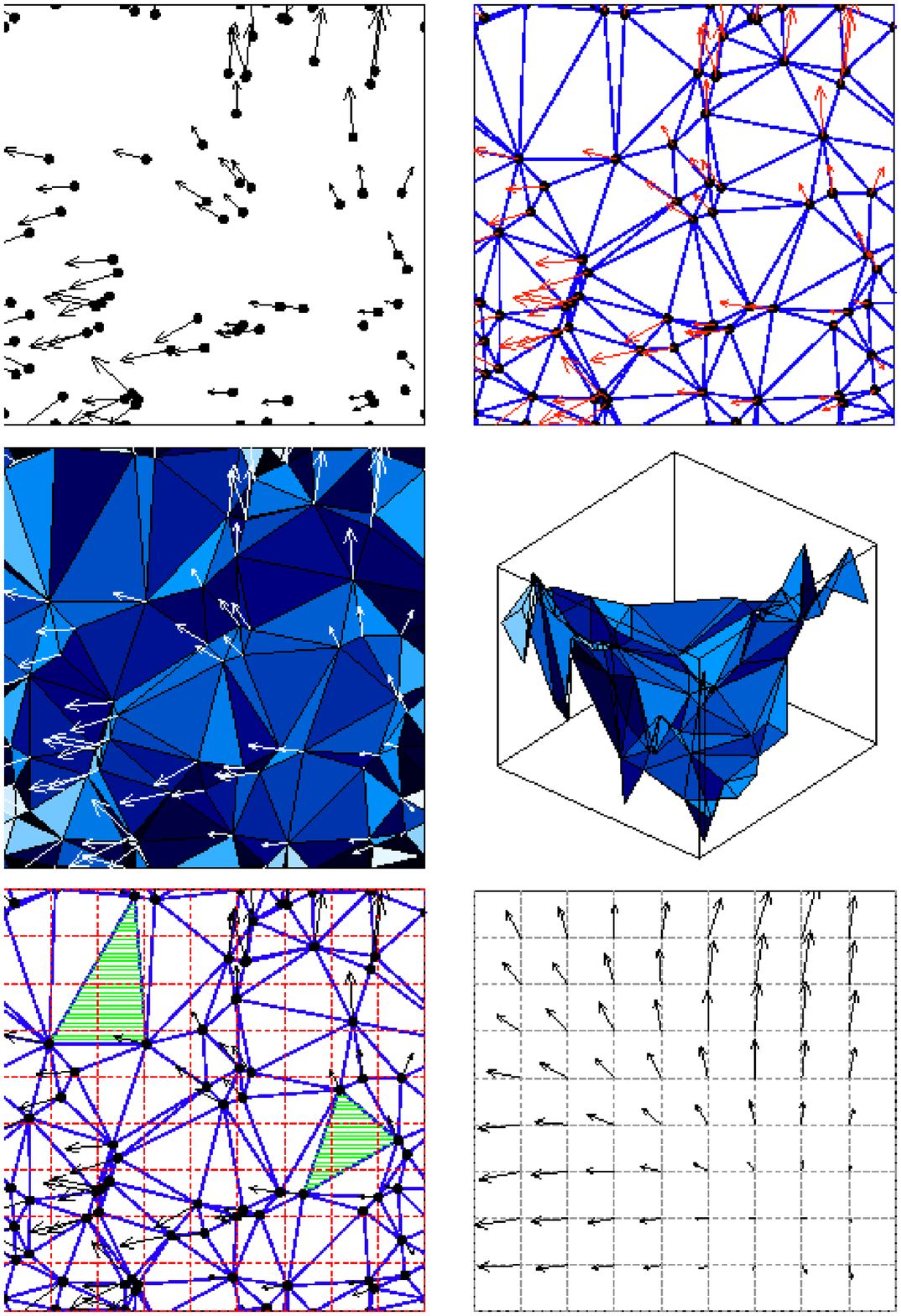}}
    \end{center}
  \end{minipage} 
\end{figure*}
\begin{figure*}
 \caption{{\small DTFE velocity interpolation procedure. The top-right
  panel presents the Delaunay triangulation of the discrete particle
  positions of the velocity field presented at the top-left panel. The
  central panels show the velocity gradient computed for each
  triangle. The colour scale corresponds to the amplitude of the
  determinant of the velocity gradient tensor. The right-hand panel
  depicts the 3D representation of the gradient surface. The height of
  each point corresponds to its velocity amplitude. The DTFE velocity
  field is estimated at the grid points indicated by the coloured grid
  of the bottom-left panel by assuming a linear variation of the
  velocity field. The bottom-right panel presents the outcome of the
  DTFE velocity interpolation procedure.}}
  \label{fig:method_dtfe_vel}
\end{figure*} 

The final product of the linearization velocity reconstruction
procedure is a galaxy catalog containing for each individual galaxy
its {\it real spatial position} and its {\it peculiar velocity}. Our
DTFE analysis of the \pscz density and velocity field is based upon
this density and linearized velocity field.


\section{The DTFE method}
\label{sec:pscz-dtfe}
The Delaunay Tessellation Field Estimator (DTFE) interpolation method
was introduced by \cite{schaapwey2000} \citep[see also][]{willemphd},
for rendering fully volume-covering and volume-weighted fields from a
discrete set of sampled field values, including the density field as
defined by the point sampling itself. DTFE involves an extension of
the interpolation procedure described by \cite{bernwey96}, who used
Delaunay tessellations for the specific purpose of estimating the
cosmic velocity divergence field, and showed the method's superior
performance with respect to conventional interpolation procedures.

\subsection{Voronoi and Delaunay tessellations}
\label{sec:delvor}
The DTFE method is based upon the {\it Voronoi and Delaunay
tessellations} of the point sample \citep[see][, and references
therein]{okabe2000}. A Delaunay tetrahedron is defined by the set of
four points whose circumscribing sphere does not contain any of the
other points in the generating set \citep{delaunay1934} (triangles in
2D). The Delaunay tessellation is intimately related to the Voronoi
tessellation of the point set, they are each others {\it dual}. The
Voronoi tessellation of a point set is the division of space into
mutually disjunct polyhedra, each polyhedron consisting of the part of
space closer to the defining point than any of the other points
\citep{voronoi1908,okabe2000}. These spatial volume-covering divisions
of space into mutually disjunct triangular (2-D) or tetrahedral (3-D)
cells adapt to the local density and geometry of the point
distribution.

Astronomical exploitation of Voronoi and Delaunay tessellations have
as yet remained rather limited, despite the fact that a great many
studies are based on observations which by their nature represent a
discretely and irregularly sampled probe of an underlying smooth
distribution. Nonetheless, in recent years the self-adaptive virtues
of Voronoi tessellations have been recognized in a gradually
increasing stream of astronomical studies. Most of these seek to use
the density sensitivity of Voronoi cell volumes and even Delaunay cell
volumes to detect density peaks like clusters of galaxies amidst a
general background of galaxies.  In astronomy, the first application
of Voronoi tessellations as density/intensity estimators was forwarded
by \cite{ebeling1993}, who sought to identify X-ray clusters as
overdensities in X-ray photon counts. More recently,
\cite{ramella2001}, \cite{marinoni2002}, \cite{kim2002} and
\cite{lopes2004} used similar algorithms to isolate clusters within
catalogues of galaxy positions produced by galaxy sky or redshift
surveys. The explicit and additional ability of the more elaborate
DTFE formalism to trace sharp density contrasts impelled
\cite{bradac2004} to compute the surface density map for a galaxy from
the projection of the DTFE volume density field. The obtained surface
density map was used to compute the gravitational lensing pattern
around the object, upon which \cite{li2006} evaluated the method in
its ability to trace higher-order singularities.

\subsection{Local spatial adaptivity}
\label{sec:nnadapt}
The point in case for its pattern tracing characteristics is provided
by the right-hand panel of Figure~\ref{fig:pscz_lin_vel} \citep[also
see fig.~4 in ][]{schaapwey06a}, showing the 2-dimensional Delaunay
triangulation for a section along the $z$-supergalactic plane through
the galaxy set which we analyze in this work, the \pscz ~catalog. The
DTFE method exploits these virtues and adapts automatically and in an
entirely natural fashion to changes in the density or the geometry of
the distribution of sampling points. Instead of involving user-defined
filters which are based upon artificial smoothing kernels the
resulting main virtue of DTFE is that it is intrinsically {\it
self-adaptive}. In essence, it involves filtering kernels which are
defined by the {\it local density and geometry} of the point process
or object distribution. The Voronoi tessellation is used to obtain
optimal {\it local} estimates of the spatial density \citep[see
sect. 8.5][]{okabe2000}, while the tetrahedra of its {\it dual}
Delaunay tessellation are used as multidimensional intervals for
linear interpolation of the field values sampled or estimated at the
location of the sample points \citep[][ch. 6]{okabe2000}.

On the basis of its interpolation characteristics DTFE is the first
order version of a wider class of tessellation-based {\it
multidimensional} and entirely {\it local} interpolation procedures,
commonly known as {\it Natural Neighbour Interpolation}
\citep[][]{watson1992,sbm1995,sukumar1998} \citep[also see][,
ch. 6]{okabe2000}. In a variety of applied sciences the concept of
natural neighbour interpolation and other advanced tessellation
applications have already found wide application. Particularly
successful and noteworthy examples may be found in computer
visualisation and surface rendering, geophysics \citep[see
e.g.][]{sbm1995,sambridge1999} and engineering mechanics
\citep{sukumar1998}.

\subsection{DTFE: density and velocity fields}
\label{sec:dtfedensvel}
A crucial aspect of the success of the Delaunay interpolation
procedure \citep{bernwey96} is that it reproduces the {\it
volume-weighted} velocity field, correcting a few fundamental biases
in estimates of higher order velocity field moments. While, often
unintentionally, most conventional interpolation schemes yield the
{\it mass-weighted} velocity, it are the {\it volume-weighted}
velocity estimates which figure in analytical expressions. This is
particularly true for velocity field perturbation analyses within the
context of gravitational instability scenarios based on primordial
Gaussian perturbations. The validity of the Delaunay (and therefore
also DTFE) velocity field interpolation scheme is perhaps most
strongly emphasized by its success in reproducing the nonlinear
velocity divergence distribution function, specifically in its
detailed agreement with second order perturbation theory
\citep{bernwey96}. The derived distribution function appears to be so
accurate that it enables the accurate determination of cosmological
parameters \citep{bernwey97}. This is even more interesting as a mildy
non-Gaussian velocity divergence distribution would enable the
breaking of the degeneracy between the cosmic matter density $\Omega$
and the bias $b$ between the matter and galaxy
distribution\citep{bernardeau94,lokas1995}.

An essential and unique step of the DTFE procedure concerns the
determination of field values at the sample points. For the velocity
field this simply involves the measured velocities at the sample
points. More complex is the issue for the density/intensity
field. Tessellation-based methods for estimating the density have been
introduced by \cite{brown1965} and \cite{ord1978}. On the condition
that the sample points are sampled proportionally to an underlying
density/intensity field, the Voronoi tessellation is used to define
optimal -- local -- density estimates \citep[][section
8.5]{okabe2000}. A minor modification was introduced
by~\cite{schaapwey2000}: in order to assure the mass-conserving nature
of the DTFE interpolation procedure the sample point density estimates
relate to the {\it contiguous} Voronoi cell (see
fig.~\ref{fig:pscz_lin_vel}).

Amongst the NN-neighbour schemes, DTFE is unique in including the
tessellation-based density estimates on the basis of the sampling
point distribution itself. By restricting itself to a linear
interpolation scheme it remains feasible to apply the method to data
sets of millions of points. The latter is essential for the viability
of the technique within a cosmological context.

\subsection{DTFE and Cosmic Structure Formation}
\label{sec:dtfestrucform}
Three major characteristics of the Megaparsec scale Universe are the
(1) the {\it weblike} spatial arrangement of galaxies and mass into
elongated filaments, sheetlike walls and dense compact clusters, (2)
the existence of large near-empty {\it void regions} and (3) the
hierarchical nature of the mass distribution, marked by substructure
over a wide range of scales and densities. According to the standard
paradigm of cosmic structure formation, the theory of gravitational
instability \citep{peebles80} this intricate spatial pattern has
emerged as a result of the gravitational growth of tiny (Gaussian)
density perturbations and the accompanying tiny velocity perturbations
in the primordial Universe. \cite{schaapwey06a} demonstrate the
ability of DTFE to succesfully reproduce and quantify these key
aspects of the nonlinear weblike cosmic matter distribution in the
Megaparsec universe:
\begin{itemize}
\item[-] The hierarchically structured matter distribution is resolved
to the smallest possible resolution scale set by the particle number
density.
\item[-] DTFE retains the morphology cq. shape of the features and
patterns in the matter distribution. The characteristic anisotropic
filamentary and planar features of the {\it Cosmic web} are fully
reproduced in the continuous DTFE density field.
\item[-] The near-empty voids in the spatial matter distribution are
reproduced optimally.  Both their flat internal density distribution
as well as their sharp outline and boundary are recovered in detail
through the interpolation characteristics of the DTFE algorithm as
well as by its tendency to suppress the shot noise in these sparsely
sampled regions \citep{willemphd}.
\end{itemize}

The relation between the cosmic density and cosmic velocity field is
an important piece of information on the dynamics of the cosmic
structure formation process. When DTFE is applied to N-body
simulations of structure formation, it does manage to succesfully
reproduce the physical and spatial correlation between cosmic density
and velocity fields for highly nonlinear structures. An early analysis
of a GIF \nbody simulation did show a few remarkable examples
\citep[see e.g.][]{willemphd}: the velocity flows in and around the
cores of high-density regions -- for as long as it concerns
single-stream laminar flows -- are traced in detail by the DTFE
density and velocity maps. Within the same simulations, the voidlike
regions were rendered as super-Hubble expanding bubbles, consistent
with our view of void dynamics \citep{icke1984, shethwey2004}.
Considerably more detailed recent
studies~\citep{schaapwey2003,emiliophd,willemphd} confirmed that DTFE
manages to trace the correspondence between velocity and density field
over a large range of scales, resolving both small and large scale
features of the velocity field.

Extrapolating this observation, and encouraged by the success of
Voronoi-based methods in identifying dark halos in N-body simulations
\citep{neyrinck2005} \cite{arad2004} used DTFE to assess the
six-dimensional phase space density distribution of dark halos in
cosmological $N$-body simulations. While a fully six-dimensional
analysis may be computationally cumbersome \citep{ascalbin05}, the
splitting of the problem into a separate spatial and velocity-space
three-dimensional tessellation may indeed hold promise for an
innovative analysis of the dynamical evolution of dark halos.

\subsection{DTFE: alternatives}
The performance of DTFE has been tested extensively with respect to
the performance of the rigid grid TSC method and the spatially
adaptive SPH method. The TSC procedure is rigid with respect to the
spatial scale as well as the shape of the mass distribution. SPH
kernels are adaptive with respect to the local density of points but
lack sensitivity to the geometry of the mass distribution.

\cite{schaapwey06a} present numerous tests of the ability of DTFE to
resolve spatial substructure and to detect anisotropic features. With
respect to spatial resolution, DTFE manages not only to resolve maps
of self-similar Soneira-Peebles models with the highest spatial
resolution but also to recover the underlying scaling indices. Both
the TSC and SPH fields fail completely in reproducing the proper
scaling properties. While TSC does not produce any scaling at all, SPH
does manage to reproduce scaling of the density field over a wide
range of spatial scales. However, it fails fully in recovering the
proper scaling indices. In addition, DTFE is also the only procedure
whose density field has an {\it autocorrelation function} which agrees
completely, down to the smallest scales, with the {\it two-point
correlation function} of the Soneira-Peebles point distribution.

Also with respect to its ability to recover weblike features in the
cosmic matter distribution the virtue of DTFE with respect to the TSC
and SPH methods becomes more evident. While the SPH procedure fares
considerably better than TSC its spherical smoothing kernel does
introduce some artefacts and deficiencies when it gets to resolving
the finest features. Defined by a fixed number of neighbours, SPH
tends to smear out the smallest structures, particularly for
anisotropically shaped ones, and results in features which occupy a
significantly larger volume than the corresponding galaxy
distribution.  However, also DTFE patterns do contain some artefacts,
of which the triangular imprint of its smoothing kernel is the most
pronounced one.

\subsection{The DTFE general reconstruction procedure}
\label{sec:dtfe_recons}
Figure~\ref{fig:method_dtfe_vel} describes the various stages of the
DTFE field reconstruction procedure. It does so with reference to
velocity field reconstruction. For a specification of the DTFE density
field procedure we refer to \cite{willemphd} \citep[also
see][]{schaapwey06a}.  In summary, it consists of the following
sequence of steps
\begin{enumerate}
\item[$\bullet$] {\bf Point sample}\\
Defining the spatial distribution of the point sample:
\begin{enumerate}
\item[+] {\it Density field}:\\ point sample needs to be a general
Poisson process of the (supposed) underlying density field, i.e.  it
needs to be an unbiased sample of the underlying density field.
\item[+] {\it General (non-density) field}:\\ no stringent
requirements upon the stochastic representativeness of the sampling
process will be necessary except that the sample volume is adequately
covered. In other words, the sample points need not form a uniform
sample of the underlying density field.
\end{enumerate}
\medskip
\item[$\bullet$] {\bf Boundary Conditions}\\ An important issue, wrt
the subsequent Delaunay tessellation computation and the
self-consistency of the DTFE density and velocity field
reconstructions, is that of the assumed boundary conditions. These
will determine the Delaunay and Voronoi cells that overlap the
boundary of the sample volume. Dependent upon the sample at hand, a
variety of options exists:\\
\begin{enumerate}
\item[+] {\it Vacuum boundary conditions:}\\ outside the sample volume
there are no points. This will lead to infinitely extended
(contiguous) Voronoi cells surrouding sample points near the
boundary. Evidently, these cells cannot be used for DTFE density field
estimates and field interpolations: the volume of the DTFE
reconstruction is smaller than that of the sample volume. \\
\item[+] {\it Periodic boundary conditions:}\\ the point sample is
supposed to be repeated periodically in boundary boxes, defining a
toroidal topology for the sample volume. The resulting Delaunay and
Voronoi tessellations are also periodic, their total volume exactly
equal to the sampel volume. While specific periodic tessellation
algorithms do exist~\citep{weygaert94}, this is not yet true for most
available routines in standard libraries. For the analysis of N-body
simulations this is the most straightforward and preferrable choice.\\
\item[+] {\it Buffer conditions:}\\ the sample volume box is
surrounded by a bufferzone filled with a synthetic point sample. The
density of the synthetic buffer point sample should be related to the
density in the nearby sample volume. The depth of the bufferzone
depends on the density of the synthetic point sample, it should be
sufficiently wide for any Delaunay or Voronoi cell related to a sample
point not to exceed the bufferzone. A clear condition for a
sufficiently deep bufferzone has been specified by
\cite{neyrinck2005}.\\ \ \\ When involving velocity field analysis,
the velocities of the buffer points should also follow the velocity
distribution of the sample and be in accordance with the continuity
equation. Relevant examples of possible choices are:\\ \itemitem{-}
{\it internal:} the analyzed sample is a subsample embedded within a
large sample volume, a sufficient number of these sample points
outside the analysis volume is taken along in the DTFE reconstruction.
\itemitem{-} {\it random cloning technique:} akin to the technique
described by \cite{yahil1991}.  \itemitem{-} {\it constrained random
field:} realizations employing the existing correlations in the field
\citep[][]{edbert87,hofmrib91,weyedb1996}.
\end{enumerate}
\medskip
\item[$\bullet$] {\bf Delaunay Tessellation}\\
Construction of the Delaunay tessellation from the point sample (see
fig.~\ref{fig:pscz_lin_vel}).  While we still use our own
Voronoi-Delaunay code~\cite{weygaert94}, at present there is a score
of efficient library routines available. Particularly noteworthy is
the \cgal initiative, a large library of computational geometry
routines\footnote{\cgal is a \texttt{C++} library of algorithms and
data structures for Computational Geometry, see \url{www.cgal.org}.}\\
\medskip
\item[$\bullet$] {\bf Field values point sample}\\
Dependent on whether it concerns the densities at the sample points or
a measured field value there are two options:
\begin{enumerate}
\item[+] {\it General (non-density) field}:\\ (Sampled) value of field
at sample point. \\
\item[+] {\it Density field}:
The density values at the sampled points are determined from the
corresponding Voronoi tessellations.  The estimate of the density at
each sample point is the normalized inverse of the volume of its {\it
contiguous} Voronoi cell ${\cal W}_i$ of each point $i$. The {\it
contiguous Voronoi cell} of a point $i$ is the union of all Delaunay
tetrahedra of which point $i$ forms one of the four vertices (see
fig.~\ref{fig:pscz_lin_vel}, 2nd frame, for an illustration). We
recognize two applicable situations:\\ \itemitem{-} {\it uniform
sampling process}: the point sample is an unbiased sample of the
underlying density field. Typical example is that of $N$-body
simulation particles. For $D$-dimensional space the density estimate
is,

\begin{equation}
{\widehat \rho}({\bf x}_i)\,=\,(1+D)\,\frac{w_i}{V({\cal W}_i)} \,.
\label{eq:densvor}
\end{equation}
\noindent with $w_i$ the weight of sample point $i$, usually we assume the same 
``mass'' for each point. \\
\itemitem{-} {\it systematic non-uniform sampling process}: sampling
density according to specified selection process.  The non-uniform
sampling process is quantified by an a priori known selection function
$\psi({\bf x})$. This situation is typical for galaxy surveys,
$\psi({\bf x})$ may encapsulate differences in sampling density
$\psi(\alpha,\delta)$ as function of sky position $(\alpha,\delta)$,
as well as the radial redshift selection function $\psi(r)$ for
magnitude- or flux-limited surveys. For $D$-dimensional space the
density estimate is ,
\begin{equation}
{\widehat \rho}({\bf x}_i)\,=\,(1+D)\,\frac{w_i}{\psi({\bf x}_i)\,V({\cal W}_i)} \,.
\label{eq:densvornu}
\end{equation}
\end{enumerate}
\medskip
\item[$\bullet$] {\bf Field Gradient}\\ Calculation of the field
 gradient estimate $\widehat{\nabla f}|_m$ in each $D$-dimensional
 Delaunay simplex $m$ ($D=3$: tetrahedron; $D=2$: triangle) by solving
 the set of linear equations for the field values at the positions of
 the $(D+1)$ tetrahedron vertices,\\
\begin{eqnarray}
\widehat{\nabla f}|_m \ \ \Longleftarrow\ \ 
\begin{cases}
f_0 \ \ \ \ f_1 \ \ \ \ f_2 \ \ \ \ f_3 \\
\ \\
{\bf r}_0 \ \ \ \ {\bf r}_1 \ \ \ \ {\bf r}_2 \ \ \ \ {\bf r}_3 \\
\end{cases}\,
\label{eq:dtfegrad1}
\end{eqnarray}
Evidently, linear interpolation for a field $f$ is only meaningful
when the field does not fluctuate strongly. Particularly relevant for
velocity field reconstructions is that there should be no orbit
crossing flows within the volume of the Delaunay cell which would
involve multiple velocity values at any one location. In other words,
DTFE velocity field analysis is only significant for {\it laminar}
flows.

Note that in the case of the sampled field being the velocity field
${\bf v}$ we may not only infer the velocity gradient in a Delaunay
tetrahedron, but also the directly related quantities such as the {\it
velocity divergence}, {\it shear} and {\it vorticity}.
\medskip

\item[$\bullet$] {\bf Interpolation}.\\ The final basic step of the
DTFE procedure is the field interpolation. The processing and
postprocessing steps involve numerous interpolation calculations, for
each of the involved locations ${\bf x}$.
\medskip
Given a location ${\bf x}$, the Delaunay tetrahedron $m$ in which it
is embedded is determined. On the basis of the field gradient
$\widehat{\nabla f}|_m$ the field value is computed by (linear)
interpolation,
\begin{equation}
{\widehat f}({\bf x})\,=\,{\widehat f}({\bf x}_{i})\,+\,{\widehat {\nabla f}} \bigl|_m \,\cdot\,({\bf x}-{\bf x}_{i}) \,.
\label{eq:fieldval}
\end{equation}
\medskip 
In principle, higher-order interpolation procedures are also possible. Two relevant 
procedures are:
\itemitem{-} Spline Interpolation
\itemitem{-} Natural Neighbour Interpolation\\
Implementation of natural neighbour interpolations is presently in progress. 

\medskip
\item[$\bullet$] {\bf Processing}.\\

Though basically of the same character for practical purposes we make
a distinction between straightforward processing steps concerning the
production of images and simple smoothing filtering operations on the
one hand, and more complex postprocessing on the other hand.  The
latter are treated in the next item. Basic to the processing steps is
the determination of field values following the interpolation
procedure(s) outlined above.\\ Straightforward ``first line'' field
operations are {\it ``Image reconstruction''} and, subsequently, {\it
``Smoothing/Filtering''}.\\
\begin{enumerate}
\item[+] {\it Image reconstruction}.\\ For a set of {\it image
points}, usually grid points, determine the {\it image value}:
formally the average field value within the corresponding gridcell.
In practice a few different strategies may be followed, dictated by
accuracy requirements. These are:\\
\itemitem{-} {\it Formal geometric approach}: integrate over the field
values within each gridcell. This implies the calculation of the
intersection of the relevant Delaunay tetrahedra and integration of
the (linearly) running field values within the
intersectio. Subsequently the integrands of each Delaunay intersection
are added and averaged over the gridcell volume.
\itemitem{-} {\it Monte Carlo approach}: approximate the integral by
taking the average over a number of (interpolated) field values probed
at randomly distributed locations within the gridcell around an {\it
image point}. Finally average over the obtained field values within a
gridcell.  
\itemitem{-} {\it Singular interpolation approach}: a reasonable and
usually satisfactory alternative to the formal geometric or Monte
Carlo approach is the shortcut to limit the field value calculation to
that at the (grid) location of the {\it image point}. This offers a
reasonable approximation for gridcells which are smaller or comparable
to that of intersecting Delaunay cells, on the condition the field
gradient within the cell(s) is not too large.
\bigskip
\item[+] {\it Smoothing} and {\it Filtering}:
\itemitem{-} Linear filtering of the field ${\widehat f}$: convolution
of the field ${\widehat f}$ with a filter function $W_s({\bf x},{\bf
y})$, usually user-specified,
   \begin{equation}
     f_s({\bf x})\,=\,\int\,{\widehat f}({\bf x'})\, W_s({\bf x'},{\bf y})\,d{\bf x'}     
   \end{equation}
\end{enumerate}
\medskip
\item[$\bullet$] {\bf Post-processing}.\\

The real potential of DTFE fields may be found in sophisticated
applications, tuned towards uncovering characteristics of the
reconstructed fields.  An important aspect of this involves the
analysis of structures in the density field. Some notable examples
are:\\
  \begin{enumerate}
    \item[+] Advanced filtering operations. Potentially interesting
applications are those based on the use of wavelets
\citep{vmartinez2005}.
    \item[+] Cluster, Filament and Wall detection by means of the {\it
Multiscale Morphology Filter} \citep{aragonmmf2007}.
    \item[+] Void identification on the basis of the {\it cosmic
watershed} algorithm~\citep{platen2007}.
    \item[+] Halo detection in N-body simulations
    \citep{neyrinck2005}.
    \item[+] The computation of 2-D surface densities for the study of
gravitational lensing \citep{bradac2004}.
  \end{enumerate} 
\medskip
In addition, DTFE enables the simultaneous and combined analysis of
density fields and other relevant physical fields. As it allows the
simultaneous determination of {\it density} and {\it velocity} fields,
it can serve as the basis for studies of the dynamics of structure
formation in the cosmos.  Its ability to detect substructure as well
as reproduce the morphology of cosmic features and objects implies
DTFE to be suited for assessing their dynamics without having to
invoke artificial filters.\\
\begin{enumerate}
  \item[+] DTFE as basis for the study of the full {\it phase-space}
structure of structures and objects. The phase-space structure dark
haloes in cosmological structure formation scenarios has been studied
by \cite{arad2004}.
\end{enumerate}
\end{enumerate}


\begin{figure} 
  \begin{center}
    \includegraphics[width=0.45\textwidth]{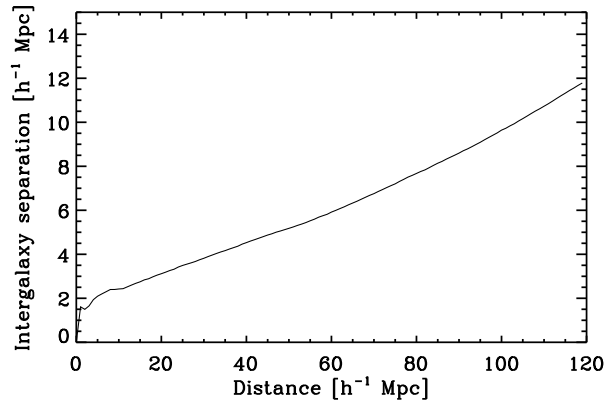}
  \end{center}
  \caption{{\small The (mean) \pscz intergalaxy distance as a function
   of distance (to the Local Group), an expression of the diminishing
   sampling density in the flux-limited \pscz sample.}}
  \label{fig:pscz_intergal}
\end{figure} 

\section{PSCz: towards the\\ 
\ \ \ \ \ DTFE density and velocity field}
Given the determination of the {\it positions} and {\it velocities} of
the \pscz galaxies in our sample by the linearization procedure of
\cite{branchini99}, the continuous volume-weighted DTFE density and
velocity field -- and the corresponding velocity divergence and shear
field -- throughout the sample volume is computed following the steps
outlined in section~\ref{sec:dtfe_recons}.

\begin{figure*}
  \begin{minipage}{\textwidth}
    \begin{center}
      \includegraphics[width=\linewidth]{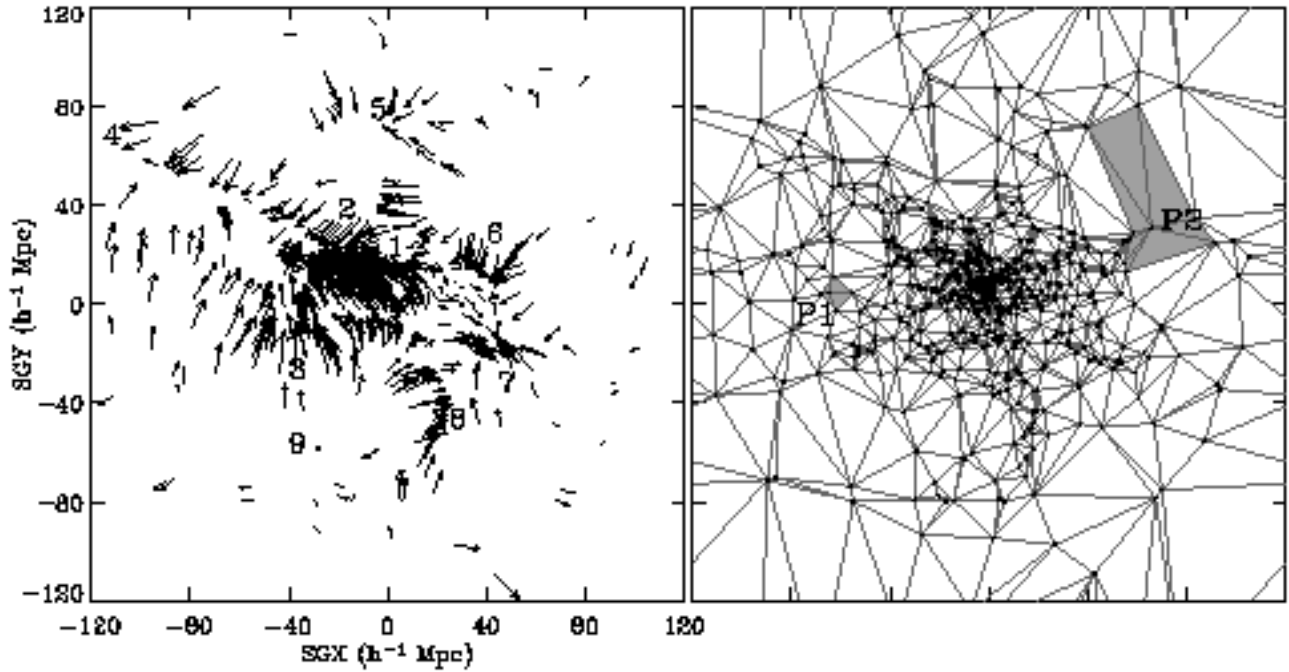}
    \end{center}
  \end{minipage} 
  \caption{{\small Modelled peculiar velocity field at the galaxy
      positions (left-hand panel).  The plot represents a slice of
      $2.5 \hmpc$ thickness centered along the $z-$supergalactic
      plane. Numbers indicate the major visible structures along the
      section. 1- Local supercluster, 2- Great attractor region, 3-
      Pavo-Indus-Telescopium complex, 4- Shapley supercluster, 5- Coma
      cluster, 6- Camelopardalis cluster, 7- Pisces-Perseus 
      supercluster, 8- Cetus wall, 9- Sculptor void.
      \citep[From][]{branchini99}. The right-hand panel shows
      the corresponding Delaunay tessellation of the galaxy distribution.
      The shadowed regions illustrate the ``contiguous Voronoi cell''
      concept for two galaxies, marked P1 \& P2.}}
  \label{fig:pscz_lin_vel}
\end{figure*} 

In our study we analyze and assess the DTFE density and velocity maps
within an inner region of radius $120 \hmpc$. It allows a more than
qualitative comparison with the analysis of \cite{branchini99} and
\cite{schmoldt1999}. Out to this radius the sample contains $10651$
galaxies. To include a sufficiently large region enclosing the main
sources gravitationally contributing to the velocity field in the
inner $120 \hmpc$ we make use of the data within a spherical volume of
radius $180 \hmpc$~\citep[][]{branchini99,emiliophd}.  Out to this
radius the \pscz sample still has sufficient structure resolution,
without shotnoise becoming dominant. The number of galaxies in this
`''\pscz-source'' sample out to this radius is $13432$.

An important factor to take into account in the interpretation and
analysis of the resulting DTFE field maps is the fact that the
inter-galaxy seperation is an increasing function of distance.  For
the \pscz sample it rises from near $\approx 2\hmpc$ within a radius
of $20 \hmpc$ to $\approx 14\hmpc$ at the outer radius of $120 \hmpc$
(see fig.~\ref{fig:pscz_intergal}). As a result the Delaunay
tessellation polyhedra will gradually grow in volume as a function of
distance.  Because it are the polyhedral cells which set the local
resolution of the interpolated fields we need to take into account
that the pure DTFE reconstruction maps have a diminishing resolution
towards larger distances. This makes postprocessing a necessity step
for a quantitative analysis and/or for an objective assessment of the
resulting density and velocity fields.

\subsection{PSCz: the Delaunay map}
The spatial positions and peculiar velocities of the galaxies in the
processed \pscz sample are depicted in the left-hand panel of
Figure~\ref{fig:pscz_lin_vel}.  The figure shows the galaxies within a
slice of $2.5 \hmpc$ thickness centered along the $z-$supergalactic
plane. The Local Group is located at the origin. Velocity vectors,
indicating the velocity component within the supergalactic plane, are
plotted at the galaxy positions. The size of the velocity vectors is
normalized with respect to the maximum velocity amplitude within the
slice. In the map the main large scale structures have been
labeled. For a description of these we refer to
section~\ref{sec:dtfepscz_cosmo}.

For reference the righthand frame depicts the Delaunay triangulation
defined by the projected positions of the sample galaxies within the
$2.5 \hmpc$ thick slice around Supergalactic Plane. By showing the
meticulous spatial adaptivity of the Delaunay tessellation, it forms a
good illustration of the rationale behind using it as an estimate for
local density. The figure emphasizes this by means of the two gray
areas, the {\it contiguous Voronoi cells} surrounding a sample
point. The inverse of these form the DTFE method's local density
estimate (see eqn.~\ref{eq:densvor}). The figure also provides a good
impression of the corresponding non-uniform spatial resolution of the
Delaunay interpolation grid.  The spatial resolution of the DTFE
interpolation grid follows the density of the sample points.

\subsection{PSCz: the DTFE interpolation}
\label{sec:dtfe_recons_pscz}
Once the spatial position ${\bf r}_n$ and velocity ${\bf v}_n$ of each
galaxy $n$ in our \pscz sample has been determined, DTFE will be
applied towards reconstructing density and velocity field.throughout
the sample volume.

\subsubsection{PSCz DTFE density values}
Evidently, the calculation of the Delaunay tessellation need to be
done only once, for both the density and velocity fields. While the
velocity values ${\bf v}_n$ at the locations ${\bf r}_n$ are part of
the input information, the density values are estimated from the
volume $V({\cal W})_n$ of the surrounding {\it contiguous Voronoi
cell} ${\cal W}_n$ around each cell $n$ (see
fig.~\ref{fig:pscz_lin_vel}). For the 3-dimensional density maps of
the magnitude-limited \pscz survey we have the density estimates
\citep{schaapwey2000}
\begin{equation}
{\widehat \rho}({\bf x}_n)\,=\,\frac{4}{\psi_{pscz}({\bf x}_n)\,V({\cal W}_n)} \,.
\label{eq:densdtfepscz}
\end{equation}
\noindent For the PSCz radial selection function $\psi_{pscz}$ we used
the expression described in \cite{branchini99}.

\subsubsection{DTFE density and velocity gradients}
The value of the density and velocity field gradient in each Delaunay
tetrahedron is directly and uniquely determined from the location
${\bf r}=(x,y,z)$ of the four points forming the Delaunay tetrahedra's
vertices, ${\bf r}_0$, ${\bf r}_1$, ${\bf r}_2$ and ${\bf r}_3$, and
the value of the estimated density and sampled velocities at each of
these locations, $({\widehat \rho}_0,{\bf v}_0)$, $({\widehat
\rho}_1,{\bf v}_1)$, $({\widehat \rho}_2,{\bf v}_2)$ and $({\widehat
\rho}_3,{\bf v}_3)$,
\begin{eqnarray}
\begin{matrix}
\widehat{\nabla \rho}|_m\\ 
\ \\
\widehat{\nabla {\bf v}}|_m
\end{matrix} 
\ \ \Longleftarrow\ \ 
\begin{cases}
{\widehat \rho}_0 \ \ \ \ {\widehat \rho}_1 \ \ \ \ {\widehat \rho}_2 \ \ \ \ {\widehat \rho}_3 \\
{\bf v}_0 \ \ \ \ {\bf v}_1 \ \ \ \ {\bf v}_2 \ \ \ \ {\bf v}_3 \\
\ \\
{\bf r}_0 \ \ \ \ {\bf r}_1 \ \ \ \ {\bf r}_2 \ \ \ \ {\bf r}_3 \\
\end{cases}\,
\label{eq:dtfegrad}
\end{eqnarray}

\noindent The four vertices of the Delaunay tetrahedron are both
necessary and sufficient for computing the entire $3 \times 3$
velocity gradient tensor $\partial v_i/\partial x_j$. Evidently, the
same holds for the density gradient $\partial \rho/\partial x_j$. We
define the matrix ${\bf A}$ is defined on the basis of the vertex
distances $(\Delta x_n,\Delta y_n, \Delta z_n)$ (n=1,2,3),
\begin{eqnarray}
\begin{matrix}
\Delta x_n \,=\,x_n-x_0\\
\Delta y_n \,=\,y_n -y_0\\
\Delta z_n \,=\,z_n -z_0\\
\end{matrix}
\quad \Longrightarrow\quad
{\bf A}\,=\,
\begin{pmatrix}
\Delta x_1&\Delta y_1&\Delta z_1\\
\ \\
\Delta x_2&\Delta y_2&\Delta z_2\\
\ \\
\Delta x_3&\Delta y_3&\Delta z_3
\end{pmatrix}
\end{eqnarray}
\noindent Similarly defining $\Delta {\bf v}_n\,\equiv\,{\bf v}_n-{\bf
v}_0\,(n=1,2,3)$ and $\Delta
{\rho}_n\,\equiv\,{\rho}_n-{\rho}_0\,(n=1,2,3)$ it is straightforward
to compute directly and simultaneously the density field gradient
$\nabla \rho|_m$ and the velocity field gradient $\nabla {\bf
v}|_m\,=\,\partial v_i/\partial x_j\,$ in Delaunay tetrahedron $m$ via
the inversion,
\begin{eqnarray}
\begin{pmatrix}
{\displaystyle \partial \rho \over \displaystyle \partial x}\\
\ \\
{\displaystyle \partial \rho \over \displaystyle \partial y}\\
\ \\
{\displaystyle \partial \rho \over \displaystyle \partial z}
\end{pmatrix}
&\,=\,&{\bf A}^{-1}\,
\begin{pmatrix}
\Delta \rho_{1}\\
\ \\
\Delta \rho_{2}\\
\ \\
\Delta \rho_{3}\\
\end{pmatrix}\,;\nonumber\\
\ \\
\begin{pmatrix}
{\displaystyle \partial v_x \over \displaystyle \partial x} & {\displaystyle \partial v_y \over \displaystyle \partial x} & 
{\displaystyle \partial v_z \over \displaystyle \partial x}\\
\ \\
{\displaystyle \partial v_x \over \displaystyle \partial y} & {\displaystyle \partial v_y \over \displaystyle \partial y} &
{\displaystyle \partial v_z \over \displaystyle \partial y} \\
\ \\
{\displaystyle \partial v_x \over \displaystyle \partial z} & {\displaystyle \partial v_y \over \displaystyle \partial z} & 
{\displaystyle \partial v_z \over \displaystyle \partial z}
\end{pmatrix}
&\,=\,&{\bf A}^{-1}\,
\begin{pmatrix}
\Delta v_{1x} & \Delta v_{1y} & \Delta v_{1z} \\ \ \\ 
\Delta v_{2x} & \Delta v_{2y} & \Delta v_{2z} \\ \ \\ 
\Delta v_{3x} & \Delta v_{3y} & \Delta v_{3z} \\ 
\end{pmatrix}\,\nonumber
\label{eq:velgrad}
\end{eqnarray}

\subsubsection{DTFE: velocity field constraints}
DTFE interpolation of velocity field values is only feasible in
regions devoid of multistream flows. As soon as there are multiple
flows -- notably in high-density cluster concentrations or in the
highest density realms of the filamentary and planar caustics in the
cosmic web -- the method breaks down and cannot be apllied.

In the study presented here this is particularly so in high-density
clusters. The complication can be circumvented by filtering the
velocities over a sufficiently large region, imposing an additional
resolution constraint on the DTFE velocity field.  Implicitly this has
actually already been accomplished in the linearization procedure of
the velocity fields preceding the DTFE processing (see
sect.~\ref{sec:pscz}).  The linearization of the input velocities
involves a kernel size of $\sqrt{5} \hmpc$
\footnote{We have used the fact that $R_{G} = R_{TH}/\sqrt{5}$
\citep{suto1990}} for the inner $50\hmpc$, for larger distances it
gradually increases with the intergalaxy distance $l(r)$ as
$l(r)/\sqrt{5} \hmpc$ (for the PSCz intergalaxy distance see
fig.~\ref{fig:pscz_intergal}). As a result, the resolution of the
velocity field is set to a lower limit of $\sqrt{5} \hmpc$. This is
sufficient to assure the viability of the DTFE velocity field
reconstructions.

\section{PSCz:\\
\ \ \ \ \ the DTFE density field}
\label{sec:dtfe-dens}
\begin{figure*} 
  \begin{minipage}{\textwidth}
    \begin{center}
      \includegraphics[width=0.85\textwidth]{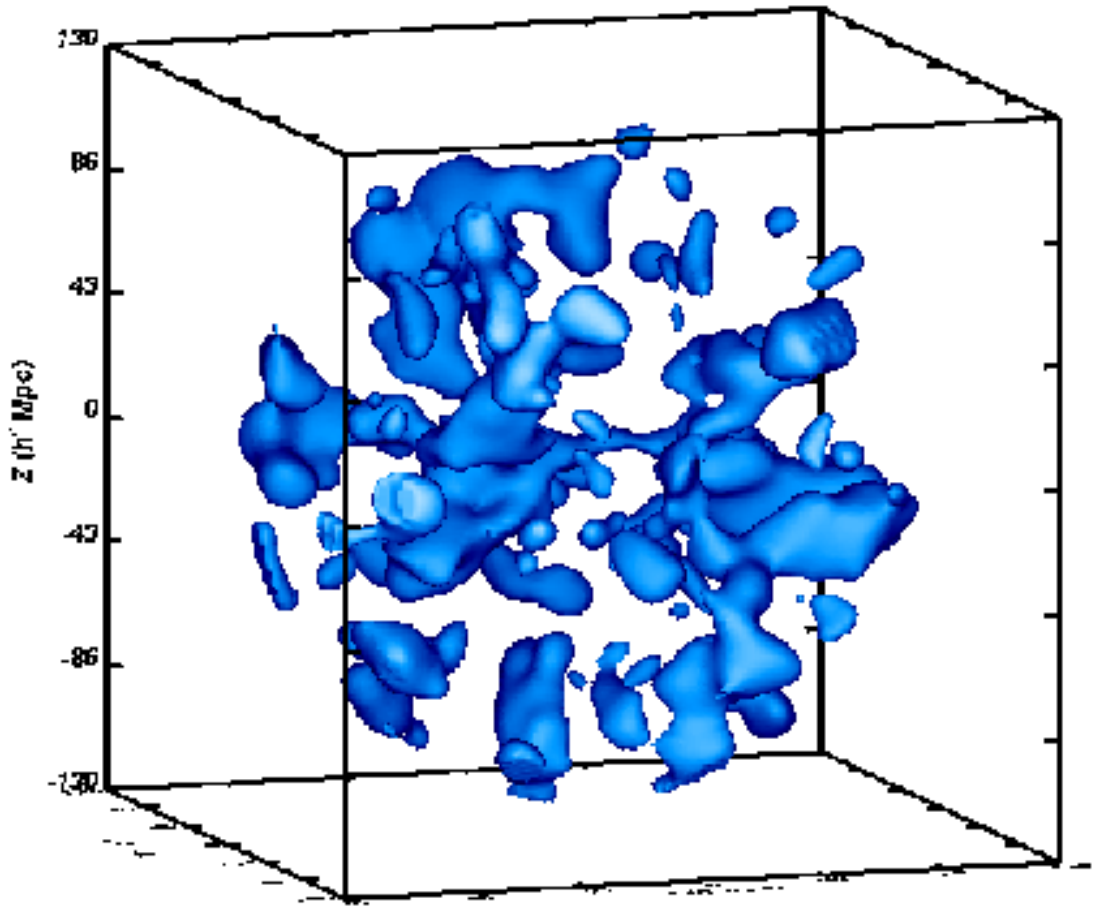}
    \end{center}
  \end{minipage} 
  \caption{{\small 3D reconstructed DTFE \pscz ~density field. The
      field has been smoothed with a Gaussian kernel of $5 \hmpc$ (for
      the effective smoothing scale see text, sect. 4.2.3). The
      isosurface represents structures at 3 times the smoothed the
      mean density. Notice the two huge density concentrations around
      the $z=0$ plane, the Pisces-Perseus and Cetus wall complex to
      the right, and the Hydra-Centaurus and Pavo-Indus-Telescopium to
      the left side. A well delineated bridge connects both
      structures, the Local Supercluster. For feature identity see the
      numbers in fig.~\ref{fig:pscz_lin_vel}.}}
  \label{fig:pscz_den_3d}
\end{figure*}

To appreciate the power of the DTFE method, the three-dimensional map
of the DTFE density field in Figure~\ref{fig:pscz_den_3d} provides a
telling illustration. For consistency between density field and the
velocity field, the density field has been smoothed with a Gaussian
kernel of $\sqrt{5} \hmpc$.  Note that for $r \largapprox 50\hmpc$ the
resolution of both density and velocity field increase proportional to
the intergalaxy distance $l(r)$: while the velocity field
linearization procedure of \cite{branchini99} explicitly involve this
as effective smoothing radius the DTFE adaptive grid resolution
automatically scales accordingly (see sect.~\ref{sec:hethomres}).  The
depicted isosurface level corresponds to structures at $3$ times the
smoothed mean density. On these scales most outstanding features
correspond to supercluster or void regions of the cosmic
web.\footnote{Notice that here, and throughout this study, the density
field $\delta$ is in effect the galaxy density field $\delta_g$. It
may be a biased reflection of the true matter density field
$\delta_m$, here parameterized by a linear bias factor $b$.}

Many familiar features are concencentrated near the Z-supergalactic
plane of the 3D density field. For identification of the various
features we show the density field in this plane by means of the
contour colour map in fig.~\ref{fig:pscz_den_vel}. The amplitude of
the corresponding density values can be inferred from the colour bar
below the map. The superimposed arrows indicate the corresponding DTFE
reconstructed velocities within the supergalactic plane (see
sect.~\ref{sec:pscz_dtfe_vel}).
 
It is interesting to compare the density map in
fig.~\ref{fig:pscz_den_vel} with the map in fig.~5 of
\cite{branchini99}. This also involves a density field, of the same
PSCz volume, smoothed with a Gaussian filter with a smoothing scale
which spatially adapts itself to the local (galaxy) density. While it
shares the superb spatial resolution with the DTFE map, the imprint of
the underlying (spherical) Gaussian filter is particularly visible in
the boundary regions of voids and superclusters.  The work by
\cite{schaapwey06a} shows that its lack of geometric adaptivtiy causes
structural artefacts near boundaries of low-density voids and
anisotropic filaments.

\subsection{The DTFE density field: Cosmography}
\label{sec:dtfepscz_cosmo}
\begin{figure*} 
  \begin{minipage}{\textwidth}
    \begin{center}
      \includegraphics[width=\linewidth]{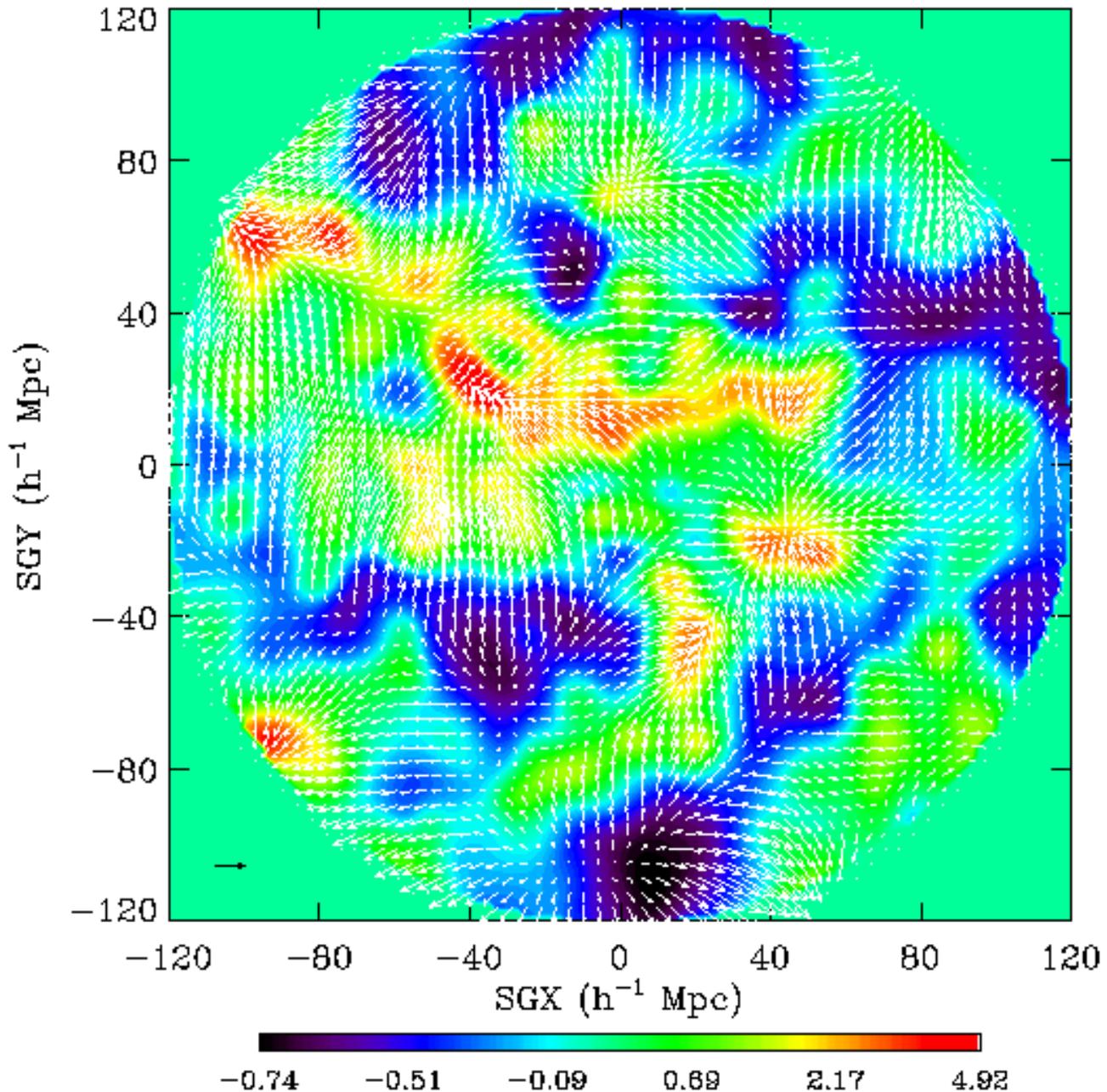}
    \end{center}
  \end{minipage} 
  \caption{DTFE density and velocity fields projected along the 
    $z-$supergalactic plane in a thin slice. The color bar indicates
    the plotted density scale. The density and velocity field have an effective 
    Gaussian smoothing radius of $R_G\sim \max{\sqrt{5},l(r)\sqrt{5}} \hmpc$, with 
    $l(r)$ the intergalaxy separation distance. Density values are 
    scaled according to the bar at the bottom of the frame, while the velocity 
    vectors are scaled such that the (black) velocity vector at the bottom left 
    corresponds to $650$ km/s.}
  \label{fig:pscz_den_vel}
\end{figure*}

The Local Group is located at the origin of the maps in
fig.~\ref{fig:pscz_lin_vel}, fig.~\ref{fig:pscz_den_3d} and in
fig.~\ref{fig:pscz_den_vel}, it is embedded in the surrounding Local
Supercluster region. On the basis of their gravitational influence on
the surroundings, manifest in the corresponding velocity field, most
of the main large-scale structures can be easily recognized in
fig.~\ref{fig:pscz_lin_vel}. For the purpose of guidance and
identification we have inserted labels in the map of the supergalactic
plane galaxy positions in fig.~\ref{fig:pscz_lin_vel}, indicating the
various large-scale structures in our neighbourhood.

High-density regions (reddish regions) as well as low density ones
(dark zones) can be easily recognized along the isodensity map in
fig.~\ref{fig:pscz_den_3d}. One of the immediate observations is that
the high-density regions, particularly the highly resolved one in the
inner region of the 3D map, tend to be flattened or elongated, a
direct consequence of the DTFE ability to trace and reproduce the
natural shape and anisotropy of structural features.

\begin{figure*}
   \vskip -1.0truecm
   \center
      \includegraphics[width=\linewidth]{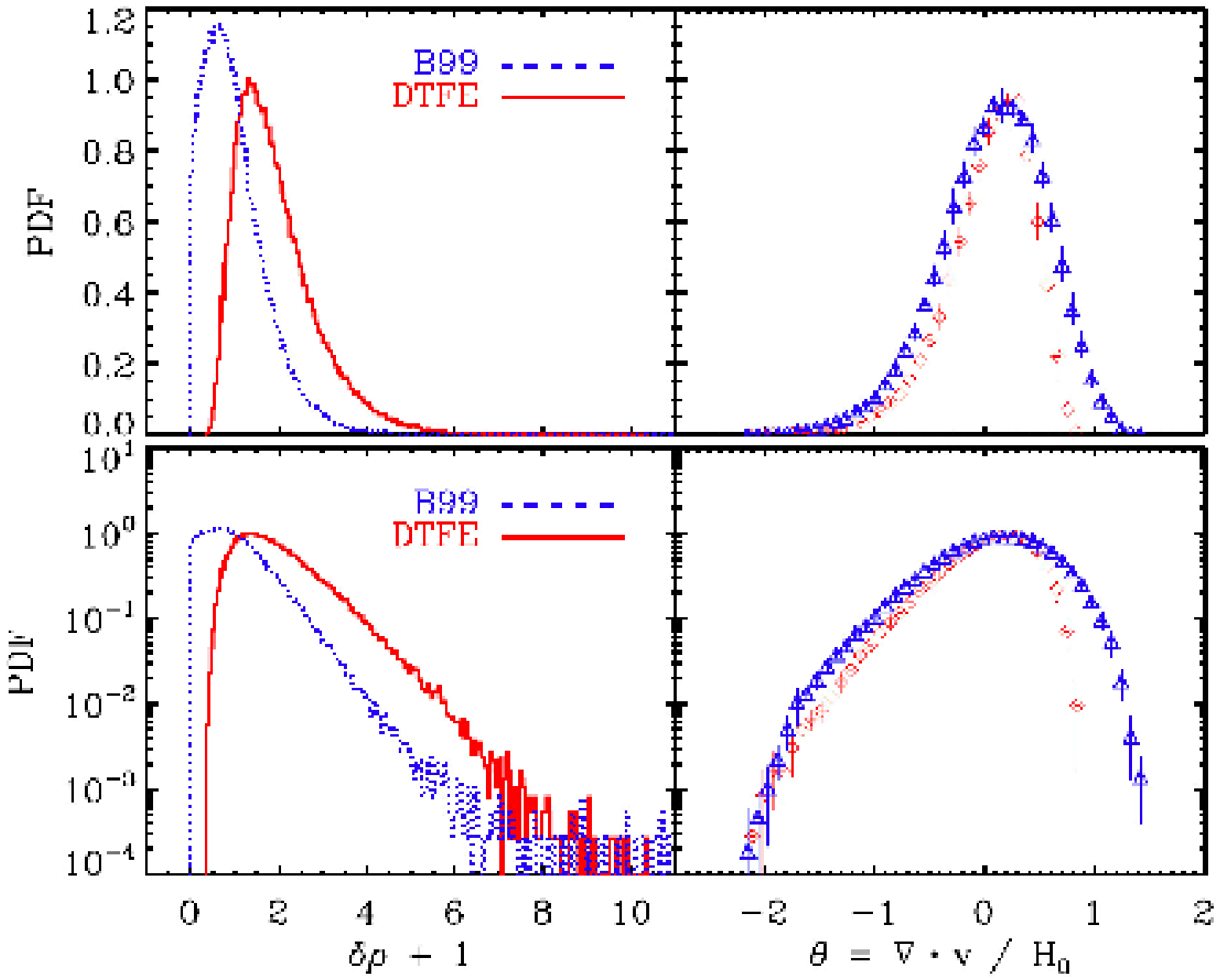}
  \caption{Density and velocity probability distribution functions
   (pdf) for PSCz. In the top row the pdfs are plotted in lin-lin, in
   the bottom row in lin-log. The lin-lin plots emphasize the level of
   non-Gaussianity of the density and velocity fields. The lin-log
   plot emphasizes prominent features in the distribution, in
   particular the sharply defined low-density edge, and facilitates
   comparison between the velocity divergence pdf and the density pdf.
   Left panels: Probability Distribution Function (PDF) of the DTFE
   density field (solid line). For comparison we have included the
   density PDF from of Branchini et al. (1999, dashed line).
   Righthand panels: the PDF of the velocity divergence (see
   sect.~\ref{sec:psczdivv}). The DTFE div v distribution is indicated
   by means of diamonds, the Branchini et al. (1999) one as triangles
   (only in the lin-log plot) as triangles. Note the good mirror-image
   correspondence between the density and div v plot for the DTFE
   fields (both lin-lin and lin-log).}.
  \label{fig:pscz_den-div_pdf}
\end{figure*} 

Two major matter concentrations along the $z=0$ plane dominate the
field (fig.~\ref{fig:pscz_den_vel}). The complex formed by the
Pavo-Indus-Telescopium supercluster and the Hydra-Centaurus
supercluster with its extension toward the Shapley concentration
dominates the left side region of the supergalactic plane (the
orientation of the figure is such that the Pavo-Indus-Telescopium
supercluster, at [SGX,SGY]$\approx[-40,-10] \hmpc$, is visible at the
front lefthand side in the 3D image). Towards the other direction of
the Local Supercluster we find a similar outstanding mass
concentration. Centrally located is the Pisces-Perseus supercluster,
extending out in the Cetus wall to the south and the Camelopardalis
cluster towards the north. The Pisces-Perseus supercluster, clearly
visible at [SGX,SGY]$\approx[45,-20] \hmpc$, and the Cetus wall, at
[SGX,SGY]$\approx[20,-40] \hmpc$ and connecting with the barely
visible Sculptor wall at [SGX,SGY]$\approx[0,-80] \hmpc$, overshadow
the other other structures.  The massive matter concentrations on
either side of the Local Supercluster are connected by a thin
filamentary bridge, passing near the origin of the map and outlining
the Local Supercluster. The filament runs from the Camelopardalis
cluster ([SGX,SGY]$\approx[45,20] \hmpc)$ towards the Shapley
supercluster ([SGX,SGY]$\approx[-120,70] \hmpc)$ and connects the
Local Supercluster with the Hydra-Centaurus Supercluster.

The Hydra-Centaurus supercluster, connecting with the Local
Supercluster on the northwest side, may be largely identified with the
Great Attractor region (GA). Also the Shapley concentration, further
out along the northwest axis, may be a major contributor to the
velocity flows in the Local Supercluster. This can be inferred from
Fig.~\ref{fig:pscz_lin_vel}, which shows that the modelled linearized
velocity field does not show any evidence for a backflow into the
Great Attractor region
\citep[e.g.][]{branchini99,rowanrobinson2000,plionis1998,basil1998}. It
might be the result of a mere artefact of the reconstruction procedure
while it did reveal itself in the map computed by means of the
alternative procedure of \cite{schmoldt1999}. Recent studies of the
dipole in the X-ray cluster distribution \citep{kocevski2006} do
indeed provide ample influence for a major dynamical influence of the
Shapley concentration, be it that the more localized 2MASS sample
appears to suggest that the tug of the GA remains the overriding
influence for the Local Supercluster \citep{erdogdu2006}.

\begin{figure*}
   \center
  \begin{minipage}{\textwidth}
    \mbox{\hskip -0.7truecm{\includegraphics[width=1.08\linewidth]{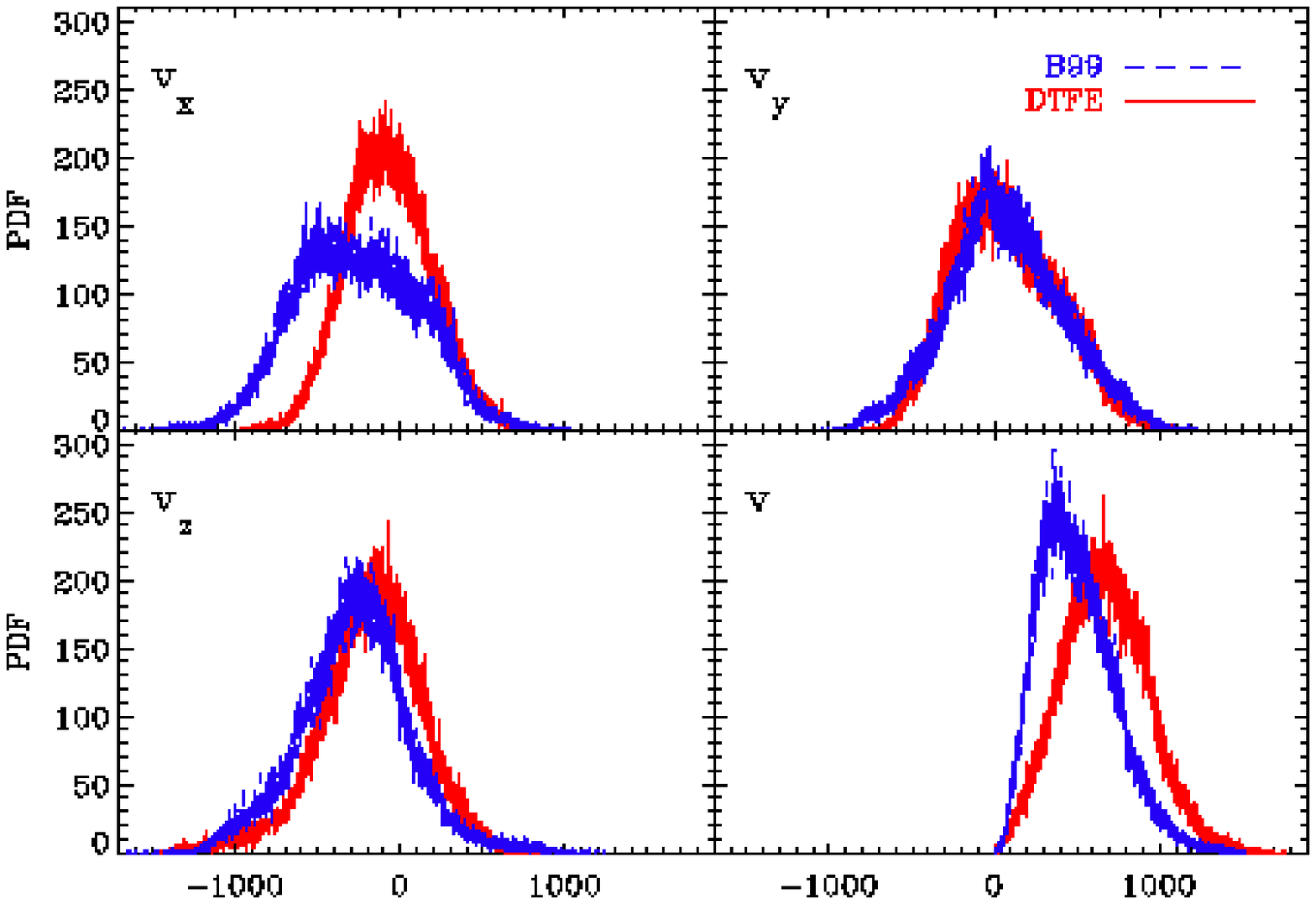}}}
   \end{minipage}
  \caption{Probability Distribution Functions for the 3 Cartesian
   velocity components and velocity amplitude. The red PDFs correspond
   to those computed by means of the DTFE technique. The blue PDFs are
   those from Branchini et al. (1999)}
\label{fig:pscz_vel_pdf}
\end{figure*} 

The Great Attractor complex and its extension towards the Shapley
configuration, the Local Supercluster and the Pisces-Perseus complex
do define a dynamical axis of a suggestive quadrupolar
configuration. The {\it tug of war} between the Great Attractor and
the Pisces-Perseus will be expressed in terms of a strong
compressional tidal force perpendicular to the axis, responsible for
the filamentary geometry and the strong velocity shear near the Local
Supercluster \citep{liljejones}, \citep[also see][]{weyedb1996}. In
all, the map of fig.~\ref{fig:pscz_den_3d} forms a wonderful
illustration of the theoretical framework of the cosmic web by
\cite{bondweb96}, in which the large scale tidal field is the web
shaping agent.

Perpendicular to the dynamical axis defined by the Great Attractor
complex, Local Supercluster and the Pisces-Perseus complex, the most
outstanding feature in our local cosmic neighbourhood is the Coma
cluster. It is visible as the only major concentration at the top of
the map, at [SGX,SGY]$\approx[0,70] \hmpc$, embedded within the Coma
wall of which the maps also allows a glimpse.

Also highly interesting are the voids visible in the DTFE density map
in the Local Universe.  These can be readily identified from the 3D
map of fig.~\ref{fig:pscz_den_3d} -- and even better from its
underdense mirror image -- as large empty (and in general roundish)
regions.  The Sculptor void, surrounded by the Pavo-Indus-Telescopium
complex and the Sculptor and Cetus walls, is one of the most salient
empty features along the supergalactic plane. Its impact on the
surrounding velocity field can be easily recognized (see
fig.~\ref{fig:pscz_lin_vel}). The Fornax void is located just at the
bottom of the plot ([SGX,SGY]$\approx[10,-110] \hmpc)$. Also the small
void located between the Coma cluster and the Hydra-Centaurus region
is clearly delineated.

\subsection{The DTFE density field: PDF}
\label{sec:dtfepscz_pdf}
Its ability to resolve small and large scale structure without loss of
spatial resolution and the ability to meaningfully interpolate the
density field in poorly sampled void regions allows DTFE to recover
the density probability distribution function, including both the low
density and high density end of the
pdf. Fig.~\ref{fig:pscz_den-div_pdf} shows the resulting distribution
function: the solid lines in the lefthand frames (top: lin-lin,
bottom: lin-log) show the pdf of the DTFE density field. The intimate
connection with the pdf of the corresponding velocity divergence
distribution will be discussed in sect.~\ref{sec:divvpdf}. By
assessing the pdf of the density and flow field restricted to the
inner $50\hmpc$ of the PSCz volume and observing it to be largely in
agreement with that for the full sample seen here we reassured
ourselves that effects resulting from the heterogeneous spatial
resolution (see sec.~\ref{sec:hethomres}) do not substantially affect
our results.

On the positive side of the PDF DTFE manages to trace the distribution
function down to pdf values of $10^{-3}$, on the negative side down to
even lower values. One direct observation is that the DTFE pdf does
not go down to density values of -1. The lin-lin plot in the lefthand
frame nicely shows the mildly non-Gaussian character of the density
field, entirely according to expectation. The lin-log plot emphasizes
the meticulous rendering of the low-density regions: on the low
density side there is a sharp cutoff at a density threshold of
$\delta=-0.8$, interestingly close to the density of a spherically
symmetric shell-crossing void in an EdS universe\footnote{The
theoretical expectation for a mature and shell-crossing void, with a
characteristic inverse tophat density profile, is an underdensity of
$\Delta \approx -1.+(1./1.7)^3 \approx -0.8$.}. It forms a powerful
confirmation of the power of DTFE to reconstruct the density
distribution in void regions.

It is particularly informative to compare the DTFE density pdf with
that of the linear density field reconstruction from
\cite{branchini99}, who used a rigid grid-based interpolation scheme
\citep[CIC,][]{hockeast}.  The dashed lines in the lin-lin and lin-log
diagram in fig.~\ref{fig:pscz_den-div_pdf} are the corresponding
pdf. Several telling differences between the DTFE and PSCz density map
of \cite{branchini99} can be identified. The gridbased reconstruction
used by the latter yields significantly lower density values in void
regions, in the order of $\delta \approx -1$. It is an expression of
the inability of the rigid gridbased interpolation to recover
meaningful density values as a result of the sparsity of sample
points: when no points are found within the grid kernel, the method
yields a zero density value. On the side of the high density values we
see that DTFE recovers systematically higher density values. This is a
consequence of the DTFE ability to trace the density field into the
most compact and/or highest density regions, without smearing these
region out as in the case of rigid kernel procedures. One may of
course argue that it would be more appropriate to compare the DTFE
results with a correspondingly smoothed CIC field. However, here we
seek to highlight the fact that DTFE does not need any such
user-specified tuning and does achieve the required filtering
automatically.

\section{PSCz:\\
\ \ \ \ \ the DTFE cosmic flow field}
\label{sec:pscz_dtfe_vel}
The DTFE reconstruction of the continuous volume-weighted DTFE
velocity field of the \pscz~catalog proceeded along the steps outlined
in section~\ref{sec:dtfe_recons}. Because the input velocities have
been linearized prior to the DTFE processing there is an implicit
limit on the resolution of the reconstructed flow field rendering
additional smoothing unnecessary. It also means that one cannot
recover features in the flow field on a scale smaller than the kernel
size ($\sqrt{5} \hmpc$ for $r \lessapprox 50 \hmpc$,
$\propto\,l(r)/sqrt{5} \hmpc$ for $r \largapprox 50 \hmpc$).

\subsubsection{the Supergalactic Plane}
Figure~\ref{fig:pscz_den_vel} presents the resulting velocity field by
means of the projected velocity vectors within the Z-supergalactic
plane, superposed upon the corresponding DTFE density contourmaps. The
length of the velocity arrows can be inferred from the arrow in the
lower lefthand corner, which corresponds to a velocity of $650$ km/s.

The processed DTFE velocity field reveals intricate details along the
whole volume.  The first impression is that of the meticulously
detailed DTFE flow field, marked by sharply and clearly defined flow
regions over the whole supergalactic plane. Large scale bulk flows,
distorted flow patterns such as shear, expansion and contraction modes
of the velocity field are clear features uncovered by our DTFE
technique. DTFE recovers clearly outlined patches marked by strong
bulk flows, regions with characteristic shear flow patterns around
anisotropically shaped supercluster complexes, radial inflow towards a
few massive clusters and, perhaps most outstanding, strong radial
outflows from the underdense void regions.

The map of fig.~\ref{fig:pscz_den_vel} shows the success of DTFE in
converting a sample of discretely sampled velocities into a sensible
volume-covering flow field. In particular its ability to interpolate
over the low-density and thus sparsely sampled regions is striking:
the voids show up as regions marked by a near-spherical outflo. By
contrast, more conventional schemes, such as TSC or SPH
\citep[see][]{schaapwey06a,emiliophd}, meet substantial problems in
defining a sensible field reconstruction in low density regions
without excessive smoothing and thus loss of resolution. At the same
time, the local nature of the DTFE interpolation guarantees a highly
resolved flow field in high density regions.

Overall, there is a tight correspondence with the large scale
structures in the underlying density distribution. While the density
field shows features down to a scale of $\sqrt{5} \hmpc$ (within the
inner $50 \hmpc$), the patterns in the flow field clearly have a
substantially larger coherence scale, nearly all in excess of $10
\hmpc$. Of course, a strong correlation between density and velocity
field is to be expected given the artificial origin of the sample
velocities as they were generated from the galaxy redshift
distribution through the linearization procedure. Strictly speaking
this concerns linear features, although it is interesting to see that
the correspondence remains almost unanimous for mildly nonlinear
supercluster and void region. The DTFE velocity flow sharply follows
the elongated ridge of the Pisces-Perseus supercluster. In addition we
find the DTFE velocity field to contain markedly sharp transition
regions between void expansion and the flows along the body of a
supercluster. We should bare in mind that this mainly concerns the
outline of features in the velocity flow, the corresponding velocity
values remain linear in character. The latter is an artefact of the
(artificial) linearization origin of the galaxy velocities in our
sample.

Massive bulk motions are particularly concentrated near and around the
massive structure extending from the Local Supercluster (center map)
towards the Great Attractor region and the Shapley concentration. The
DTFE map nicely renders this pronounced bulk flow towards the
Hydra-Centaurus region and shows that it dominates the general motions
at our Local Group and Local Supercluster.  The most massive and
coherent bulk flows in the supergalactic plane appear to be connected
to the Sculptor void and the connected void regions (towards the
lefthand side of the figure). They are the manifestation of the
combination of gravitational attraction by the heavy matter
concentration of the Pavo-Indus-Telescopium complex, the more distant
``Hydra-Centaurus-Shapley ridge'', and the effective push by the
Sculptor void region. Conspicuous shear flows can be recognized along
the ridge defined by the Cetus wall towards the Pisces-Perseus
supercluster ([SGX,SGY]$\approx[20,-40] \hmpc$. A similar strong shear
flow is seen along the extension of the Hydra-Centaurus supercluster
towards the Shapley concentration.

The influence of the Coma cluster is beautifully outlined by the
strong and near perfect radial infall of the surrounding matter,
visible at the top-centre of figure~\ref{fig:pscz_den_vel}. Also the
velocity field near the Perseus cluster, in the Pisces-Perseus
supercluster region, does contain a strong radial inflow component.

Perhaps most outstanding are the radial outflow patterns in and around
voids.  The intrinsic suppression of shot noise effects through the
adaptive spatial interpolation procedure of DTFE highlights these
important components of the Megaparsec flow field and emphasizes the
dynamical role of voids in organizing the matter distribution of the
large scale Cosmic Web.

\subsection{Velocity Field: PDF}
We have also compared the PDFs of the Cartesian velocity components
and the total velocity amplitude. The red PDFs in
figure~\ref{fig:pscz_vel_pdf} represent those computed by means of the
DTFE technique, while the blue ones are those from \cite{branchini99}.

The DTFE velocity pdfs have a healthy Gaussian appearance, as we might
have expected for a near linear velocity field. In this respect it is
interesting to notice the stark differences between the DTFE
distribution functions and those of the velocity field of
\cite{branchini99}. In particular the velocity component in the
$x$-direction shows a marked deviation. We traced the differences back
to the low density regions, where rigid gridbased methods have
considerable difficulty in defining reasonable density values.  On the
other hand we also note significant differences in the high density
regions. Gridbased methods are unable to resolve these and tend to
average the velocities in these regions over the volume of a
gridcell. This is reflected in the pdf of the velocity amplitude: the
DTFE pdf testifies of considerably higher velocities in the DTFE
velocity field. Note that this may yield important implications for
the analysis of large scale bulk flows within the sample's volume,
conventional methods may yield unjustifiably biased values.


\begin{figure*} 
  \begin{minipage}{\textwidth}
    \begin{center}
      \includegraphics[width=\linewidth]{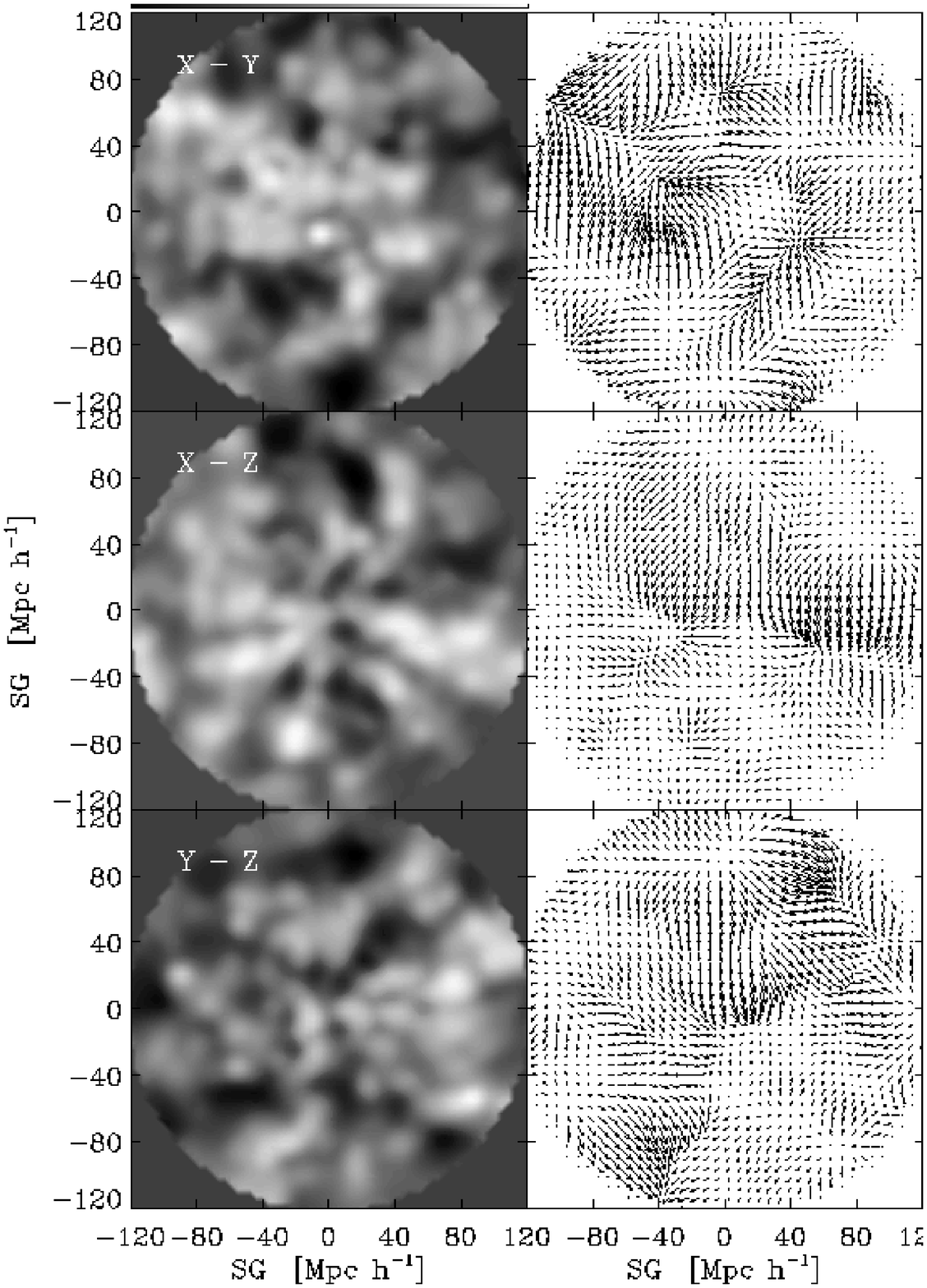}
    \end{center}
  \end{minipage} 
\end{figure*} 
\begin{figure*} 
  \begin{minipage}{\textwidth}
    \begin{center}
      \includegraphics[width=\linewidth]{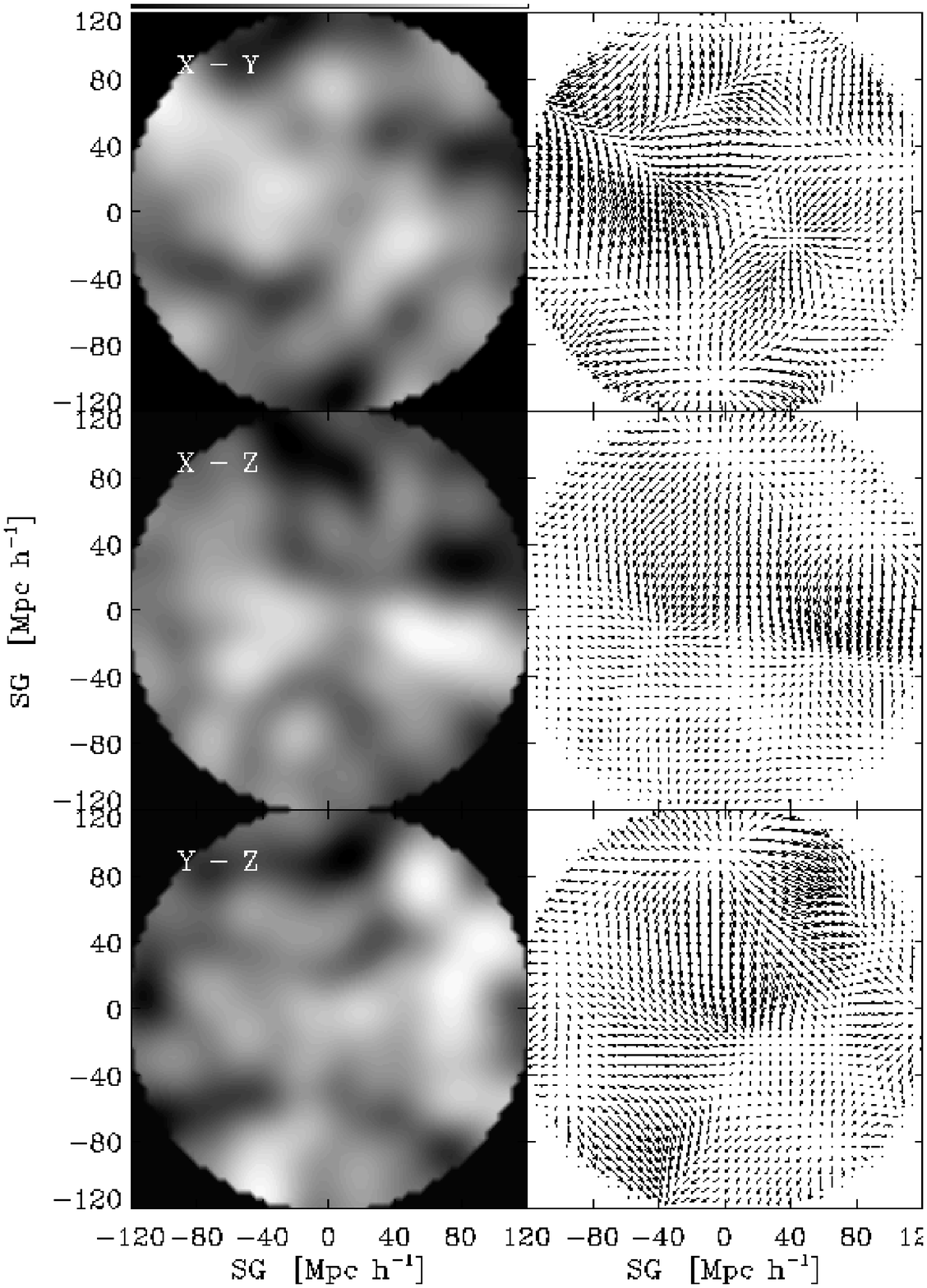}
    \end{center}
  \end{minipage} 
\end{figure*} 
\begin{figure*} 
  \caption{The DTFE density (lefthand) and velocity vector (righthand)
    maps of the PSCz sample in three mutually perpendicular central
    supergalactic planes: the X-Y, X-Z and Y-Z plane (from top to
    bottom). The density and velocity field is smoothed on a scale of
    $3.8 \hmpc$. This implies a uniform resolution up to a radius of
    $30\hmpc$ around the centre, with a gradually diminishing
    resolution towards the outer edge of the sample volume. The
    greyscale bar at the top lefthand indicates the density
    values. The velocity vectors are scaled such that the vector atop
    the figure corresponds to $650$ km/s.}
  \label{fig:pscz_den_vel_r3.8_xyz}
\end{figure*}
\begin{figure*} 
  \caption{The DTFE density (lefthand) and velocity vector (righthand)
    maps in three mutually perpendicular central supergalactic planes:
    the X-Y, X-Z and Y-Z plane (from top to bottom). The density and
    velocity field is smoothed on a scale of $12 \hmpc$, involving a
    uniform resolution over the sample volume. The greyscale bar at
    the top lefthand indicates the density values. The velocity
    vectors are scaled such that the vector atop the figure
    corresponds to $650$ km/s.}
  \label{fig:pscz_den_vel_r12_xyz}
\end{figure*}

\subsubsection{the Spatial Flow Field}
It remains a rather challenging task to visualize a three-dimensional
flow field. We chose to show the spatial structure of the DTFE
rendered velocity flow within the \pscz~volume within the three
mutually perpendicular central slices through the centre of the
supergalactic plane (ie. the Local Group's location). The combination
of the velocity flow within the $2.5\hmpc$ thick slices along the
supergalactic X-Y, the X-Z and the Y-Z supergalactic plane provides a
reasonable impression of the spatial flow field. To understand its
connection with the underlying matter distribution we also depict the
density distribution within the same planes.

To appreciate the structure of the \pscz velocity flow and to obtain
an impression of the density and velocity field contributions on
different scales we have included the complementary set of maps in
fig~\ref{fig:pscz_den_vel_r3.8_xyz} and
fig.~\ref{fig:pscz_den_vel_r12_xyz}. By means of a set of $3 \times 2$
frames, the latter shows the density and velocity field filtered on a
scale of $R_G=12\hmpc$, the former on $R_G=3.8\hmpc$. The density
field in the X-Y, X-Z and Y-Z central slices is represented by
greyscale maps in three consecutive rows of the lefthand column, the
greyscale values in all 3 frames are scaled according to the values in
the bar atop the figure. The corresponding velocity vector maps are
shown in the righthand column.  The velocity vectors in the three
planes are normalized such that the reference velocity vector depicted
atop the three frames corresponds to $650\,\hbox{\rm km/s}$. The view
on a scale of $R_G=12\hmpc$ is that of the large (linear) scale
contributions to the flow field and identifies the agents of these
large scale flows. The considerably more detailed map of
fig~\ref{fig:pscz_den_vel_r3.8_xyz} has the intention of identifying
the imprint of mildly nonlinear structural features such as
superclusters and voids on the corresponding velocity flow.

The combination of fig~\ref{fig:pscz_den_vel_r3.8_xyz} and
fig.~\ref{fig:pscz_den_vel_r12_xyz} is a potentially rich source for
the study of the structural dynamics in the local Universe. Important
for the present study is that it emphasizes three key issues of the
DTFE velocity field analysis:
\begin{enumerate}
\item[$\bullet$] the spatial adaptivity of DTFE allows the resolution
of the velocity flow in and around nonlinear features in the density
field.
\item[$\bullet$] the spatial adaptivity of the DTFE kernel to the
density of sample points implies a dependence of the spatial
resolution of the (raw) DTFE velocity field on the sampling
density. As a result high resolution maps do involve a rather
inheterogeneous spatial resolution over the sample volume.
\item[$\bullet$] the DTFE velocity field has a more homogeneous
resolution than the corresponding density field. This is a result of
the larger scale character of cosmic velocity fields.
\end{enumerate}

\section{DTFE maps: Spatial Resolution}
\label{sec:mapresolution}
The cosmic matter distribution and cosmic velocity flows include
contributions over a wide range of scales. One of the virtues of DTFE
is its large dynamic range, a result of its adaptivity
\citep[see][]{schaapwey06a}. In principal we are therefore equipped
for an assessment of the structure of the produced DTFE PSCz density
and velocity maps in terms of their structure at various
scales. However, before doing so we need to pay attention to the fact
that for the flux-limited sample involved the DTFE maps do not
necessarily involve a homogeneous resolution.

\subsection{Heterogeneous vs. Homogeneous resolution}
\label{sec:hethomres}
The cosmographical presentation of the pure DTFE density and velocity
map in fig.~\ref{fig:pscz_den_vel} in sec.~\ref{sec:dtfepscz_cosmo}
does not include the issue of its rather non-uniform, radially
declining, spatial resolution. This issue becomes prominent in the
comparison between the maps of fig.~\ref{fig:pscz_den_vel_r12_xyz} and
those of fig.~\ref{fig:pscz_den_vel_r3.8_xyz}. While the latter have a
uniform but low spatial resolution of $R_G=12\hmpc$, the higher
resolution maps filtered on a scale $R_G=3.8\hmpc$ do lack such
uniform resolution.

By its nature, DTFE resolves regions of a higher sampling density on a
finer scale, as can be readily inferred from the structure of the
Delaunay grid in fig.~\ref{fig:pscz_lin_vel}. As a result the spatial
resolution diminishes when the intergalaxy separation in the sample
increases. The increase of the \pscz~intergalaxy separations as a
function of distance (fig.~\ref{fig:pscz_intergal}) thus provies an
order of magnitude estimate of the effective resolution scale as
function of radial distance from the Local Group\footnote{Note that
the resolution scale is not exactly a function of radial distance, it
is entirely determined by the local sampling density. On average this
does increase as function of radial distance. See
fig.~\ref{fig:pscz_lin_vel}}.

A resolution scale of $R_G\sim 3.8\hmpc$ corresponds to a radial depth
of $\sim 30\hmpc$ while $R_G\sim 12\hmpc$ is reached at a distance of
$120\hmpc$. By implication the maps in
fig.~\ref{fig:pscz_den_vel_r3.8_xyz} are uniform in the inner
$30\hmpc$, but they are marked by a gradually diminishing spatial
resolution at larger distances. It are in particular the density maps
which show the gradually fading resolution of the $3.8\hmpc$ maps most
clearly. By contrast, the $R_G=12\hmpc$ maps show a uniformly resolved
view of the structure and flow in the Local Universe. A scale of
$R_G=12\hmpc$ corresponds to the sample intergalaxy distance at the
outer edge of the PSCz sample.

For a proper quantitative analysis and assessment of the DTFE density
and velocity field reconstructions it is of the utmost importance to
take into account the gradually diminishing spatial resolution of the
raw DTFE maps. In this respect we also need to take into account that
cosmic density and velocity fields are composed of contributions from
a range of scales. By implicitly filtering out small-scale
contributions in regions with a lower sampling density DTFE will
include a smaller range of spatial scales contributing to a field
reconstruction. On average, they will therefore correspond to regions
with a lower amplitude, a subtle point of importance in any
quantitative analysis. A comparison of DTFE density and velocity maps
in the PSCz volume shows that the latter seems to be hardly affected
by this effect, a manifestation of the fact that the cosmic velocity
field is dominated by larger scale modes than the density field
\citep{peebles80}.

\subsubsection{The Large Scale Local Universe:\\
\ \ \ \ \ \ \ \ \ the $12 \hmpc$ maps} 
In the large scale $R_G=12\hmpc$ density and velocity field of
fig.~\ref{fig:pscz_den_vel_r12_xyz} we recognize nearly all features
of the previous discussion on the structure of the supergalactic plane
(sect.~\ref{sec:dtfepscz_cosmo}). The X-Y supergalactic plane clearly
involves a local matter concentration: while the X-Y plane represents
a more or less even distribution of in particular high-density
features, we observe a concentration of high-density regions towards
the corresponding Z=0 axis in the X-Z density map and a more moderate
density distribution with some occasional high-density patches in the
Y-Z map. One of the most conspicuous mass concentrations in the latter
is the Coma cluster, at around [SGY,SGZ]$\approx[70,0] \hmpc$,
embedded within a massive ``Great Wall'' complex extending along the
$Z$ axis. It clearly emphasizes the moderate density values near our
own LOcal Supercluster, lying in between the massive Hydra
Centaurus-Shapley concentration and the Pisces-Perseus concentration
in the X-Y plane. The only massive feature which had not been traced
in the X-Y map is the concentration near [SGY,SGZ]$\approx[-40,-100]
\hmpc$.

The presence of massive matter concentrations in the $R_G=12\hmpc$ map
induces strongly correlated patterns in the flow field. The X-Z map is
dominated by a strong and rather coherent flow towards the Z$=0$ (X-Y)
supergalactic plane.  Interesting is to see that this flow is tied in
with an almost equally prominent outflow from the low-density regions
towards the upper righthand of the maps. A similar pattern can be
recognized in the Y-Z map, where near the centre we observe an equally
massive flow towards the supergalactic Z=0 plane. It appears to be
coupled to a truely massive radial outflow from the underdensity near
[SGY,SGZ]$\approx[0,80] \hmpc$. If anything, this demonstrates the key
role of low density regions in the dynamics of the local Universe, and
in the formation of cosmic structure in general.

Within the X-Y supergalactic plane we recognize the familiar flow
patterns discussed in the previous section,
sect.~\ref{sec:dtfepscz_cosmo}. The coherent bulk flow along the Local
Supercluster-Centaurus-Shapley axis. It is coupled to coherent towards
the Pisces-Perseus region, via a saddle point near the Local Group
location: the dominant shear flow in the Local Universe. Radial
outflow low-density regions are somewhat less prominent, the outflow
from the local Sculptor void appears to be merged into a large scale
coherent bulk flow.

\subsubsection{Flow Field scale dependence:\\
\ \ \ \ \ \ \ \ \ $3.8\hmpc$ vs. $12.0\hmpc$} 
The impact of DTFE is particularly emphasized by the comparison of the
medium resolution $3.8\hmpc$ maps and the $12.0\hmpc$ maps. The more
detailed density map of the maps in
fig.~\ref{fig:pscz_den_vel_r3.8_xyz} reveals considerably more detail
in both density and velocity maps.

The structure in the X-Y density map has already been discussed in
detail in sect.~\ref{sec:dtfepscz_cosmo}. For the $R_G=3.8\hmpc$
density there are two major points of notice in comparison to the
$R_G=12\hmpc$ maps.  The first is the gradually decreasing density
values towards the outer regions: most prominent features are
concentrated within the inner $30-40\hmpc$ radius. This is a direct
consequence of the decreasing spatial resolution of the non-uniform
$R_G=3.8\hmpc$ maps (see sect.~\ref{sec:mapresolution}).  Another
feature of interest is the more prominent presence of (deeper) void
regions wrt. the corresponding $R_G=12\hmpc$ map. This is in
particular true for the Y-Z plane.

The $R_G=3.8\hmpc$ velocity field stands out by considerably more
sharply defined and outlined features. All features that were
recognized in the $R_G=12\hmpc$ velocity field can also be identified,
be it they are much more sharply defined. Moreover, all over the
volume these sharp velocity features are clearly correlated with
features in the density field. The mildly nonlinear nature of
supercluster complexes is reflected in pronounced bulk and shear
flows. In the velocity map these often go along with remarkably
sharply defined transition regions where the flow changes abruptly as
it encounters a flow from another direction.  Also void outflows
appear to be much more prominent than the weakly positive divergence
regions in the large scale map. The DTFE void outflows have the
appearance of superhubble bubbles, and relate very well to the deep
nonlinear potential wells of true voids
\citep[see][]{schaapwey2007}. In addition, we can recognize
small-scale nonlinear features that were not seen in the $R_G=12\hmpc$
velocity map. The radial infall towards Coma and the radial outflow
out of the Sculptor void are telling examples.

Both aspects strongly underline the ability of DTFE to trace velocity
flows into nonlinear regions. This forms a strong argument for DTFE's
potential in the analysis of the dynamics of mildly nonlinear
structures, such as filamentary or wall-shaped superclusters and
voids. The DTFE analysis of Virgo CDM simulations by
\cite{schaapwey2007} does indeed provide quantitative evidence for
this ability of DTFE. Only when the nonlinear evolution of a structure
has proceeded towards a higher nonlinear stage, marked by multistream
regions and shell crossing displacements, the linear interpolation
scheme of DTFE will break down.


\begin{figure*} 
  \begin{minipage}{\textwidth}
    \begin{center}
      \includegraphics[width=\linewidth]{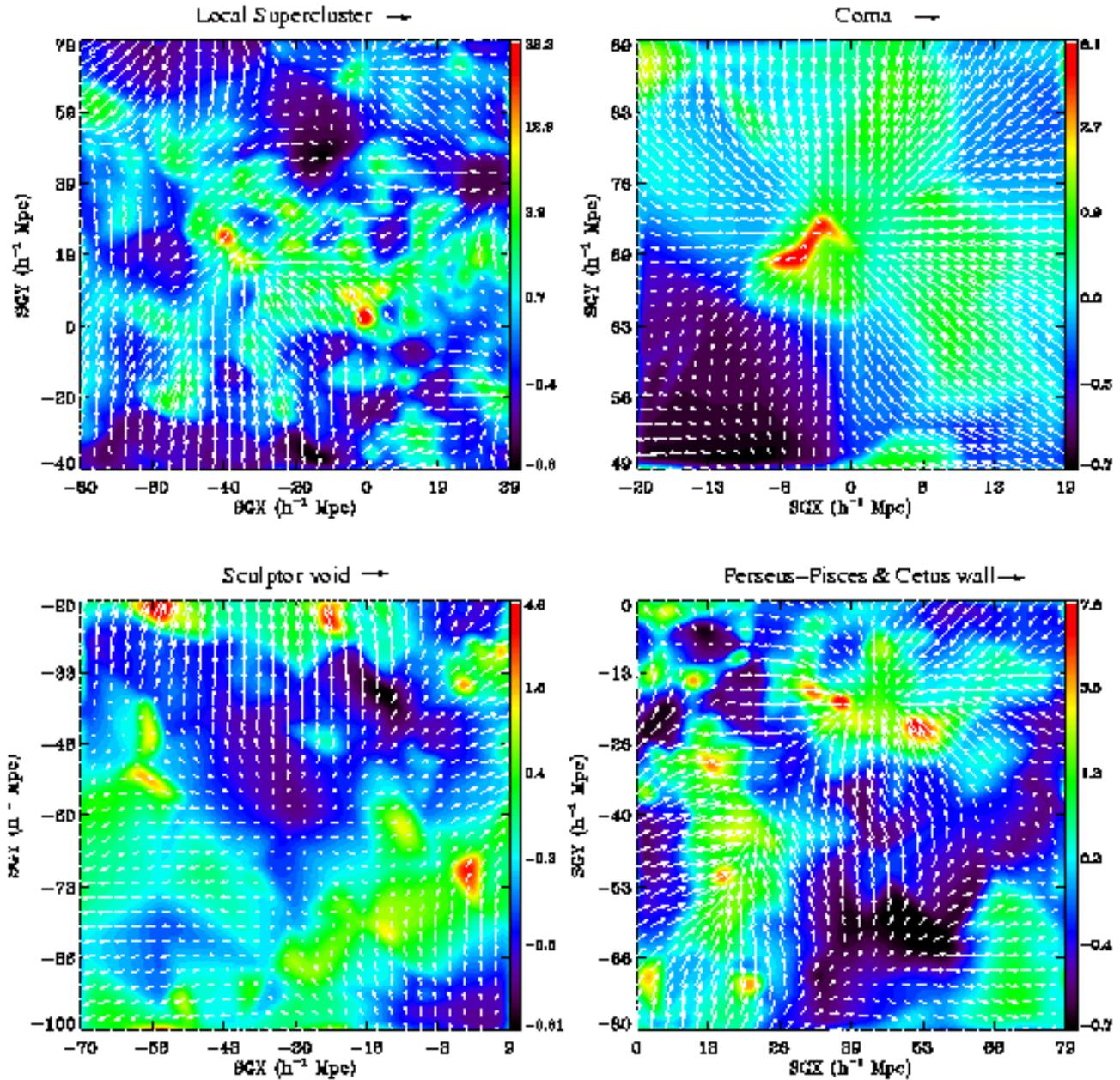}
    \end{center}
  \end{minipage}
  \caption{Density and velocity zooms for four different regions along
    the supergalactic plane indicated by the labels at the top of each
    frame. The density field has been convolved with a Gaussian kernel
    of $1 \hmpc$ for a better impression of such field. Note that the
    density colour contour values in each of the frames are chosen
    differently, their scaling reflected in the color bars at the
    right side of each frame. The velocity vectors at the top of each
    frame, at the righthand side of the frame's name, correspond to a
    velocity of $650$ km/s.}
  \label{fig:pscz_zoom}
\end{figure*} 

\section{Individual Cosmic Structures}
\label{sec:pscz_zooms}
The ability of the DTFE method to resolve and identify both small and
large scale features, on the condition of being sufficiently sampled,
makes it into a highly promising basis for the development of
structure detection algorithms.

To illustrate the variety of structures that may be identified by DTFE
we zoom into four different regions of the DTFE PSCz map, each of a
different character. The four regions are the Local Supercluster, the
Coma cluster, the Sculptor void and the Pisces-Perseus \& Cetus
supercluster filament.  The four corresponding panels in
Figure~\ref{fig:pscz_zoom} show the density field and corresponding
velocity field vectors. To ascertain optimal resolution the density
field has been smoothed with a Gaussian kernel of a mere $1 \hmpc$. It
is important to realize that as a result of the irregular spatial
resolution this does not necessarily mean that all structures of that
size are revealed. Note that the physical size of the corresponding
panels varies with the spatial extent of the depicted structure.

\subsection{the Local Supercluster}
The top-left panel shows the complex velocity field in and around the
Local Supercluster.  The Virgo cluster is located just above the Local
Group. Particularly prominent is the presence of the Great Attractor
region and its dominant influence on the velocity field. At the
location of the Local Group ([SGY,SGZ]$\approx[0,0] \hmpc)$ we find a
large bulk flow, part of a large scale shear flow towards the Great
Attractor region. The shear patterns is a result of the exerted
gravitational influence of the Great Attractor and Pisces-Perseus
supercluster citep[e.g. see][]{emiliophd}.

\subsection{the Pisces-Perseus supercluster}
The bottom-righthand frame focusses on the Pisces-Perseus supercluster
and the Cetus wall.  The prominent flow along the Pisces-Perseus chain
towards the Cetus wall is a clear indication of the dynamical
connection between these two structures. Their gravitational influence
can be traced along the whole zoomed region. The velocity field around
the underdense region located at [SGX,SGY]$\approx[45,-60] \hmpc$ is
strongly distorted by these two massive structures. Note the presence
of the shear pattern near the top-left corner, near the location of
the Local Group.

\subsection{the Coma Cluster}
The top-right hand panel shows how the Coma cluster, embedded within
the Coma wall, distorts the velocity field in its surroundings. The
DTFE reconstruction clearly depicts the almost isotropic infall into
the Coma cluster. The slight offset is an artefact of our
implementation of the DTFE interpolation towards the image grid
locations. Coma is embedded within the Coma wall. The shear pattern
visible near the bottom-left is the result of the opposing forces
between the Coma region and the Local Supercluster region.

\subsection{the Sculptor Void}
The Sculptor void and surroundings is the subject of the bottom
lefthand frame. The lowest measured DTFE density contrast value
(smoothed at $1 \hmpc$) in this region is $\approx -0.78$ at the
deepest of the void. At the smoothed scale of $\sqrt{5}\hmpc$ the DTFE
density threshold is $-0.74$, in agreement with the reported value by
\cite{plionis2002} of $-0.69$.  The velocity field of this almost
``empty'' expanding region is distorted by the surrounding matter
distribution. Small, yet detectable, distortions delineate the
boundary of this void with the surrounding matter distribution,
cq. the Sculptor wall.

\section{The DTFE \pscz~velocity divergence}
\label{sec:psczdivv}
A direct spin-off of the DTFE velocity field analysis is that it
implicitly comes along with the velocity field divergence, shear and
vorticity. In the outline of the DTFE method, in
sect.~\ref{sec:dtfe_recons}, we described how the DTFE method
determines the gradient $\widehat{\nabla f}|_m$ of the sample field
$f$ in each Delaunay simplex $m$ (eqn.~\ref{eq:dtfegrad}). For the
three spatial components of the velocity field, $(v_x,v_y,v_z)$, this
translates into a velocity gradient counting nine components $\partial
v_i/\partial x_j$ (eqn.~\ref{eq:velgrad}). Within the context of DTFE
each velocity gradient $\partial v_i/\partial x_j$ has a uniform value
within each Delaunay tetrahedron. By implication, the reconstructed
DTFE velocity {\it gradient} field is not continuous.

\begin{figure*} 
  \center
  \begin{minipage}{\textwidth}
    \mbox{\hskip 3.5truecm\includegraphics[width=0.7\linewidth,angle=90.0]{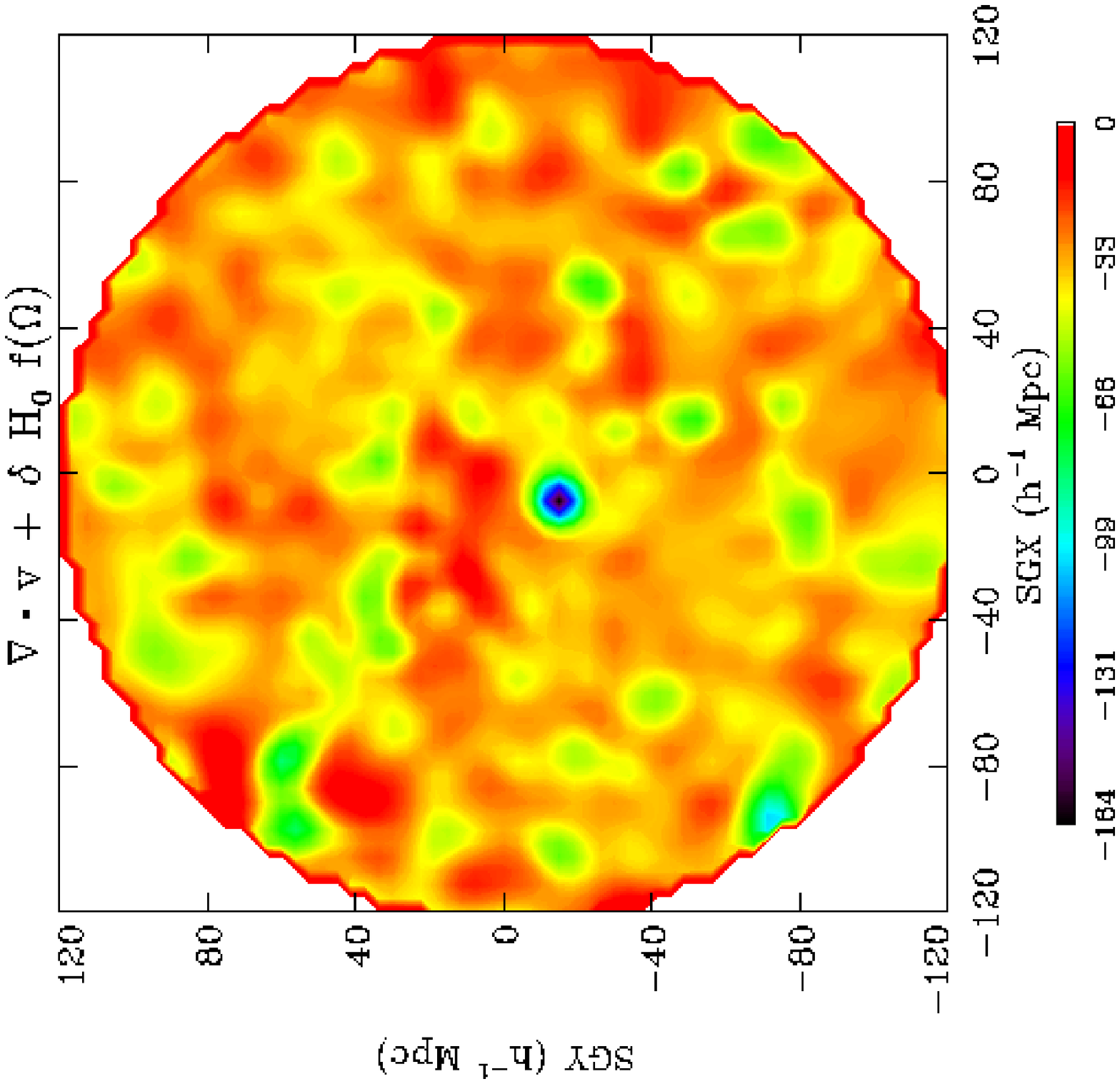}}
    \vskip -1.5truecm
    \mbox{\hskip 3.5truecm\includegraphics[width=0.7\linewidth,angle=90.0]{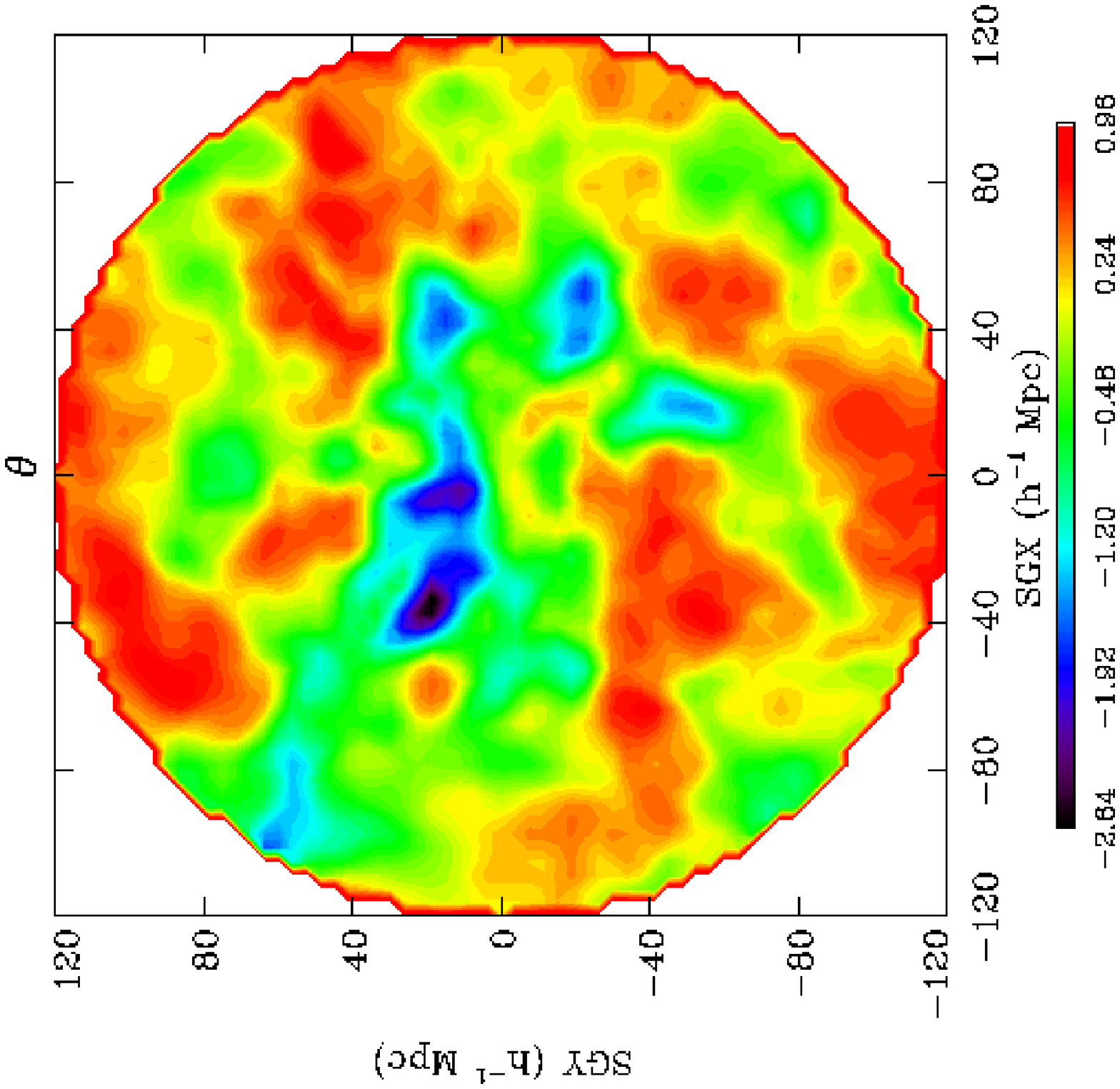}}
   \end{minipage}
\end{figure*} 
\begin{figure*} 
  \begin{minipage}{\textwidth}
    \begin{center}
      \includegraphics[width=\linewidth]{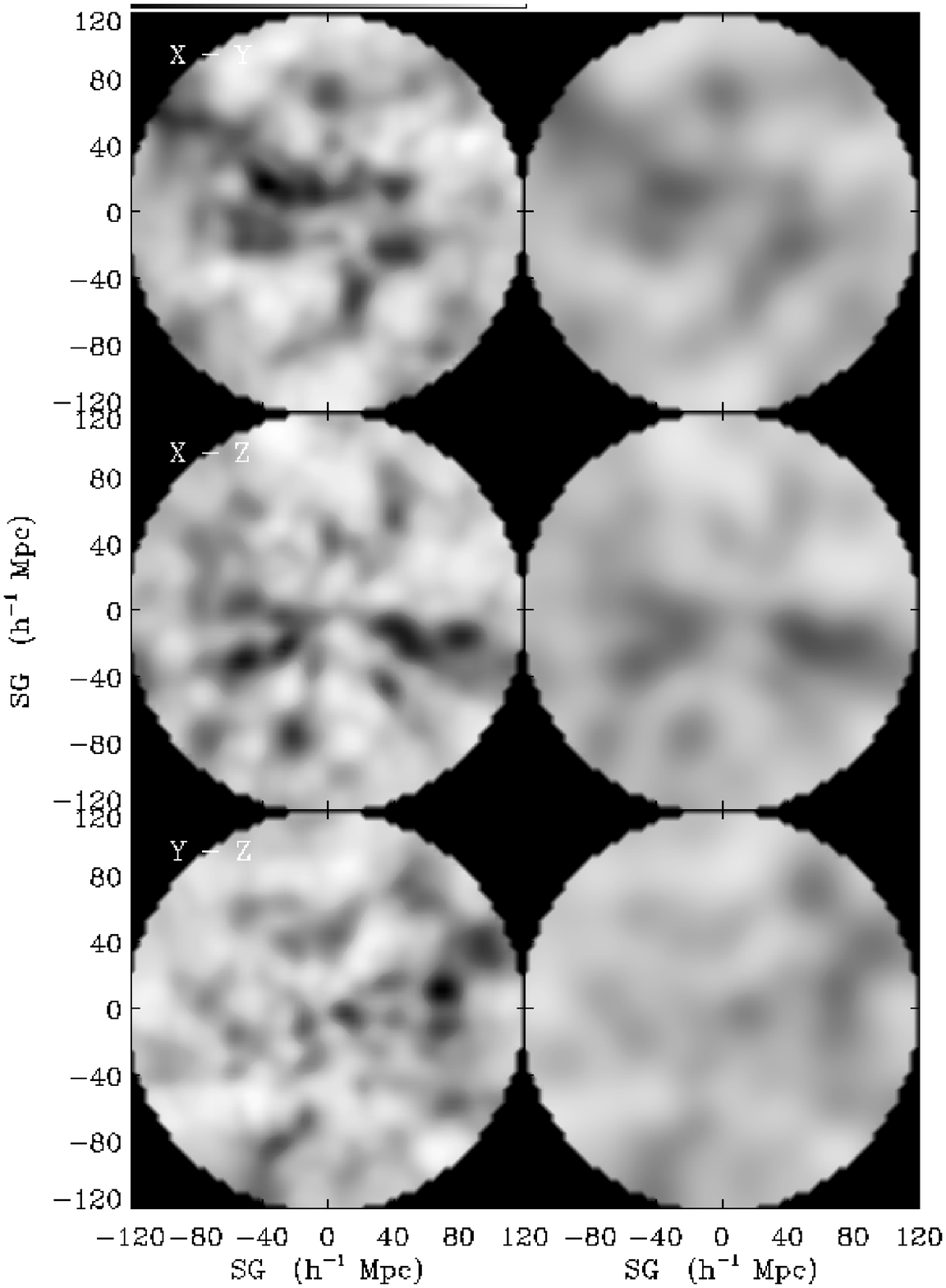}
    \end{center}
  \end{minipage} 
\end{figure*} 
\begin{figure*}
   \caption{Lefthand: DTFE velocity-divergence field, on a smoothing
    scale of $R_G=\sqrt{5}\hmpc$ projected along the Z-supergalactic
    plane. The thin slice corresponds to the one presented in
    Fig.~\ref{fig:pscz_den_vel}. The velocity-divergence is in units
    of the Hubble parameter $H_0$. The color bar represents the scale
    of the plotted velocity-divergence. Righthand: The map, along the
    Z-supergalactic plane, of the difference between the DTFE
    velocity-divergence field and the linear theory prediction for the
    corresponding DTFE density field. The color bar indicates the
    corresponding field values, in units of $\kmsmpc$.}
\label{fig:pscz_div}
\end{figure*}
\begin{figure*}
  \caption{The DTFE velocity-divergence field on a smoothing scale of
   $3.8\hmpc$ (lefthand column) and $10 \hmpc$ (righthand column). The
   maps in the consecutive rows concern the three mutually
   perpendicular central supergalactic planes: the X-Y, X-Z and Y-Z
   plane (from top to bottom). Note that the map on a scale of
   $3.8\hmpc$ is uniform only out to a radius of $30\hmpc$, with a
   gradually diminishing resolution towards the outer edge of the
   sample volume. Also compare the images with the density fields in
   fig.~fig.~\ref{fig:pscz_den_vel_r3.8_xyz} and
   \ref{fig:pscz_den_vel_r12_xyz}, they are approximately the
   negatives of the corresponding density fields. The greyscale values
   are in units of the Hubble parameter $H_0$, with their values given
   in the bar at the top of the figure.}
  \label{fig:pscz_divv_3_10_xyz}
\end{figure*}

\subsection{Velocity divergence, shear and vorticity}
From the 9 velocity gradient components $\partial v_i / \partial x_j$
we can directly determine the three velocity deformation modes, the
velocity divergence $\nabla \cdot {\bf v}$, the shear $\sigma_{ij}$
and the vorticity ${\bf \omega}$,
\begin{eqnarray}
\nabla \cdot {\bf v}&\,=\,&
\left({\displaystyle \partial v_x \over \displaystyle \partial x} +
{\displaystyle \partial v_y \over \displaystyle \partial y} +
{\displaystyle \partial v_z \over \displaystyle \partial z}\right)\,,\nonumber\\
\ \nonumber\\
\sigma_{ij}&\,=\,&
{1 \over 2}\left\{ 
{\displaystyle \partial v_i \over \displaystyle \partial x_j} +
{\displaystyle \partial v_j \over \displaystyle \partial x_i}
\right\}\,-\, {1 \over 3} \,(\nabla\cdot{\bf v})\,\delta_{ij} \,,\\
\ \nonumber\\
\omega_{ij}&\,=\,&
{1 \over 2}\left\{
{\displaystyle \partial v_i \over \displaystyle \partial x_j} -
{\displaystyle \partial v_j \over \displaystyle \partial x_i}
\right\}\,.\nonumber
\label{eq:vgradcomp}
\end{eqnarray}
\noindent where ${\bf \omega}=\nabla \times {\bf v}=\epsilon^{kij}
\omega_{ij}$ (and $\epsilon^{kij}$ is the completely antisymmetric
tensor). In the theory of gravitational instability, there will be no
vorticity contribution as long as there has not been shell crossing
(ie. in the linear and mildly nonlinear regime). For the evolution of
density perturbations, the velocity divergence is an important
velocity field component from the early linear regime onward. As
nonlinearity sets in the deforming character of accompnaying shear
flows become increasingly decisive in shaping the cosmic matter
distribution. It is in particular via the Zel'dovich formalism
\citep{zeldovich70} and the {\it Cosmic Web} theory by Bond and
collaborators that we have come to appreciate the key role of tides
and velocity shear in the formation of the observed weblike Megaparsec
matter distribution \citep{bondweb96}.
 
The relationship between the cosmic density field on the one hand and
the velocity divergence and shear fields on the other hand contains
substantial information on the dynamics and formation of structure in
the Universe. Within the context of the corresponding DTFE analysis it
is important to realize that there is an important and implicit
discrepancy: the DTFE density and velocity fields are entirely
continuous. The velocity divergence fields consist of discontinuous
patches: the Delaunay tetrahedra, each with a constant value of
$\nabla \cdot {\bf v}$.  This renders a perfect 1-1 correspondence
between the raw DTFE density and $\nabla \cdot {\bf v}$ fields is
unfeasible. Filtering over the scale of a Delaunay simplex is usually
necessary to restore the physically expected relations.

In the following subsections we present the maps of the velocity
divergence field and the velocity shear field in the same central
supergalactic planes as in figs.~\ref{fig:pscz_den_vel} and
~\ref{fig:pscz_den_vel_r3.8_xyz}. By smoothing the velocity
divergence, velocity shear and vorticity fields with the Gaussian
kernel $R_G=\sqrt{5}\hmpc$, akin to the density map in
fig.~\ref{fig:pscz_den_vel} and fig.~\ref{fig:pscz_den_vel_r12_xyz},
we restore a sense of continuity to the resulting maps.

\subsection{The DTFE velocity divergence map}
The DTFE normalized velocity divergence estimate $\widehat \theta$ is
the sum of the trace of the DTFE velocity-gradient components,
\begin{equation}
{\widehat \theta}\,\equiv\,{\displaystyle \widehat{\nabla \cdot {\bf v}} \over \displaystyle H_0}\,=
{\displaystyle 1 \over \displaystyle H_0}\,\left({\widehat{\frac{\partial v_x}{\partial x}}} +
  {\widehat{\frac{\partial v_y}{\partial y}}} + {\widehat{\frac{\partial v_z}{\partial z}}}\right) \,.
  \label{eq:dtfe_div_dtfe}
\end{equation}
\noindent with $H_0$ the Hubble constant \footnote{We have adopted a
value of $h=0.7$ ($H_0 \equiv 100 ~h \kmsmpc$)}. In
Figure~\ref{fig:pscz_div} (lefthand) we have plotted ${\widehat
\theta}$.

The expanding and contracting modes of the velocity field are clearly
delineated.  Expanding regions, corresponding with or surrounding
large and deep underdense voids, are identified as those with red to
yellow tones marking positive divergence modes. The Sculptor and
Fornax voids and the underdense regions around the Coma cluster can be
immediately recognized. Other conspicuous expanding regions are that
near the Camelopardalis cluster, located at [SGX,SGY]$\approx[40,45]
\hmpc$, and the one at the right of the Cetus wall at
[SGX,SGY]$\approx[50,-55] \hmpc$.

The regions with a negative divergence are contracting, matter is
falling in along one or more directions. The strongest contractions,
represented by the blue tones, are related to the peaks in the density
field which can be immediately identified with the most massive
structures located along this slice. The Hydra-Centaurus supercluster
is the most prominent region of infall. Easily recognizable are all
clusters such as the Virgo, Camelopardalis and Coma cluster, the
Pisces-Perseus supercluster, the Cetus wall, the
Pavo-Indus-Telescopium complex, and partly also the Shapley
concentration. These high-contrast blue regions are embedded within
regions with more moderate infall motions. These regions, with green
contour values, have densities slightly in excess of the mean. They
outline a percolating region with a roughly filamentary shape.

In the interpretation of the velocity divergence map it is crucial to
take account of an important artefact: high amplitudes of the velocity
divergence are seemingly concentrated in the inner region of the PSCz
volume. This can be understood on the same grounds as a similar
tendency for the density field, the lack of small-scale power in the
outer regions as a result of the diminishing sampling density. A
correction for this effect is only possible on the basis of a
cosmological model assumption and does not fall within the context of
this study.

\subsection{The density-velocity divergence relation}
\label{sec:den-divv}
\noindent The velocity divergence and the density contrast are related
via the continuity equation \citep{peebles80}. In the linear regime
this is a strictly linear one-to-one relation,
\begin{equation}
{\displaystyle 1 \over \displaystyle H}\,\nabla \cdot {\bf v} ({\bf x,t})\,=\,- {\displaystyle f(\Omega_m) 
\over \displaystyle b}\,a(t)\,\delta({\bf x},t)\,,
\label{eq:divvdenlin}
\end{equation}
\noindent linking the (galaxy) density perturbation field $\delta$ to
the peculiar velocity field ${\bf v}$ via the factor $f(\Omega_m)$
\citep[see][]{peebles80}. In this we take into account that our
discussion concerns the density field as sampled by galaxies, whose
distribution is supposed to be linearly biased with respect to the
underlying matter density. There remains a 1-1 relation between
velocity divergence and density into the mildly nonlinear regime (see
eqn.~\ref{eq:divvdennlin}), which explains why the map of the velocity
divergence (\ref{fig:pscz_div}) is an almost near perfect negative
image of the density map (see fig.~\ref{fig:pscz_den_vel}).

\begin{figure}
    \begin{center}
      \includegraphics[width=8.5truecm]{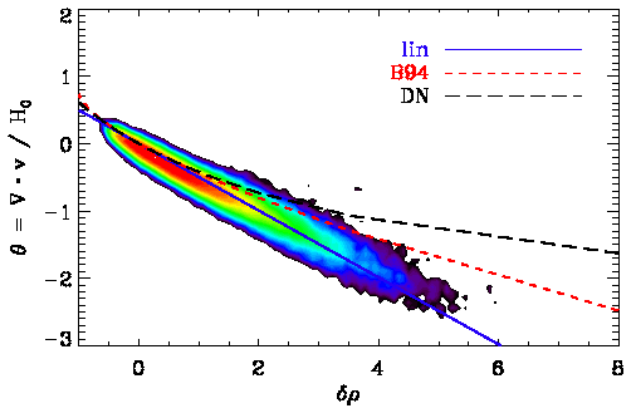}
    \end{center}
  \caption{{\small Scatter plot of density versus velocity divergence,
   in terms of a contour plot of the density of scatter points. The
   solid line indicates the linear density-velocity divergence
   relation. Two approximate nonlinear relations are also indicated,
   the short-dashed line depicts the approximation of Bernardeau
   (1992), the long-dashed one the approximation by Nusser {\it
   et~al.}(1991).}}
  \label{fig:pscz_den-divv}
\end{figure} 

The approximate validity of the linear divergence-density relation
(\ref{eq:divvdenlin}), on a scale of $R_G=\sqrt{5}\hmpc$, over the
PSCz volume can be appreciated from the contour map in the righthand
frame of fig.~\ref{fig:pscz_div}. The map, in the same supergalactic
X-Y plane as in the lefthand frame, shows the difference between the
divergence term on the lefthand side of eqn.~\ref{eq:divvdenlin} and
the density term on the righthand side, in units of
${\kmsmpc}$. Evidently, over the entire volume the residual map is
near uniform, with the exception of a ``hotspot'' near
[SGX,SGY]$\approx[-10,-10] \hmpc$. If anything, we see a slight
tendency for the differences to increase outward. This is most likely
to be understood from a different reaction of the velocity field to
the diminishing sampling density than the density field.

The residual map's fig.~\ref{fig:pscz_div} impression of near
linearity is confirmed by the scatter plot between velocity divergence
$(\nabla \cdot {\bf v})$ and (galaxy) density $\delta$ in
Figure~\ref{fig:pscz_den-divv}. The scatter point density in the
figure is shown by means of a contour map of the scatter point
density.

The scatter plot is compared with three different
relations. Superimposed as a solid line is the linear relation,
$\nabla \cdot v = -H_0 \beta \delta$. The near perfect linearity of
the $(\nabla\cdot {\bf v})$-$\delta$ relation reflects the origin of
the velocity estimates on the basis of a linearization
process. Deviations from linearity, be it minor, are observed only for
the lowest and highest density values.

Even in the quasi-linear and mildly nonlinear regime the one-to-one
correspondance between velocity divergence and density remains intact,
be it that it involves higher order terms \citep[see][for an extensive
review]{bernardeau2002}. Within the context of Eulerian perturbation
theory \cite{bernardeau92} (B) derived an accurate 2nd order
approximation form the relation between the divergence and the density
perturbation $\delta({\bf x})$. \cite{nusser1991} (N) derived a
similar quasi-linear approximation within the context of the
Lagrangian Zel'dovich approximation. According to these approximate
nonlinear relations,
\begin{eqnarray}
{\displaystyle 1 \over \displaystyle H}\,\nabla \cdot {\bf v} ({\bf x})\,=\,
\begin{cases}
{\displaystyle 3 \over \displaystyle 2} f(\Omega_m) \left[1-(1+\delta({\bf x}))^{2/3}\right]\,\qquad\hfill\hfill \hbox{\rm (B)}\\
\ \\
-f(\Omega_m)\,{\displaystyle \delta({\bf x}) \over \displaystyle 1 + 0.18\,\delta({\bf x})}\,\qquad\hfill\hfill \hbox{\rm (N)}\\
\end{cases}
\label{eq:divvdennlin}
\end{eqnarray}
\noindent for a Universe with Hubble parameter $H(t)$ and matter
density parameter $\Omega_m$. For comparison we have also included
these approximate relations for mildly nonlinear density and velocity
field in fig.~\ref{fig:pscz_den-divv} (B: the short-dashed line, N:
the long-dashed line). These relations are clearly too pronounced for
the velocity divergence map of fig.~\ref{fig:pscz_div}.

\begin{figure*} 
  \center
  \begin{minipage}{\textwidth}
    \includegraphics[width=\linewidth]{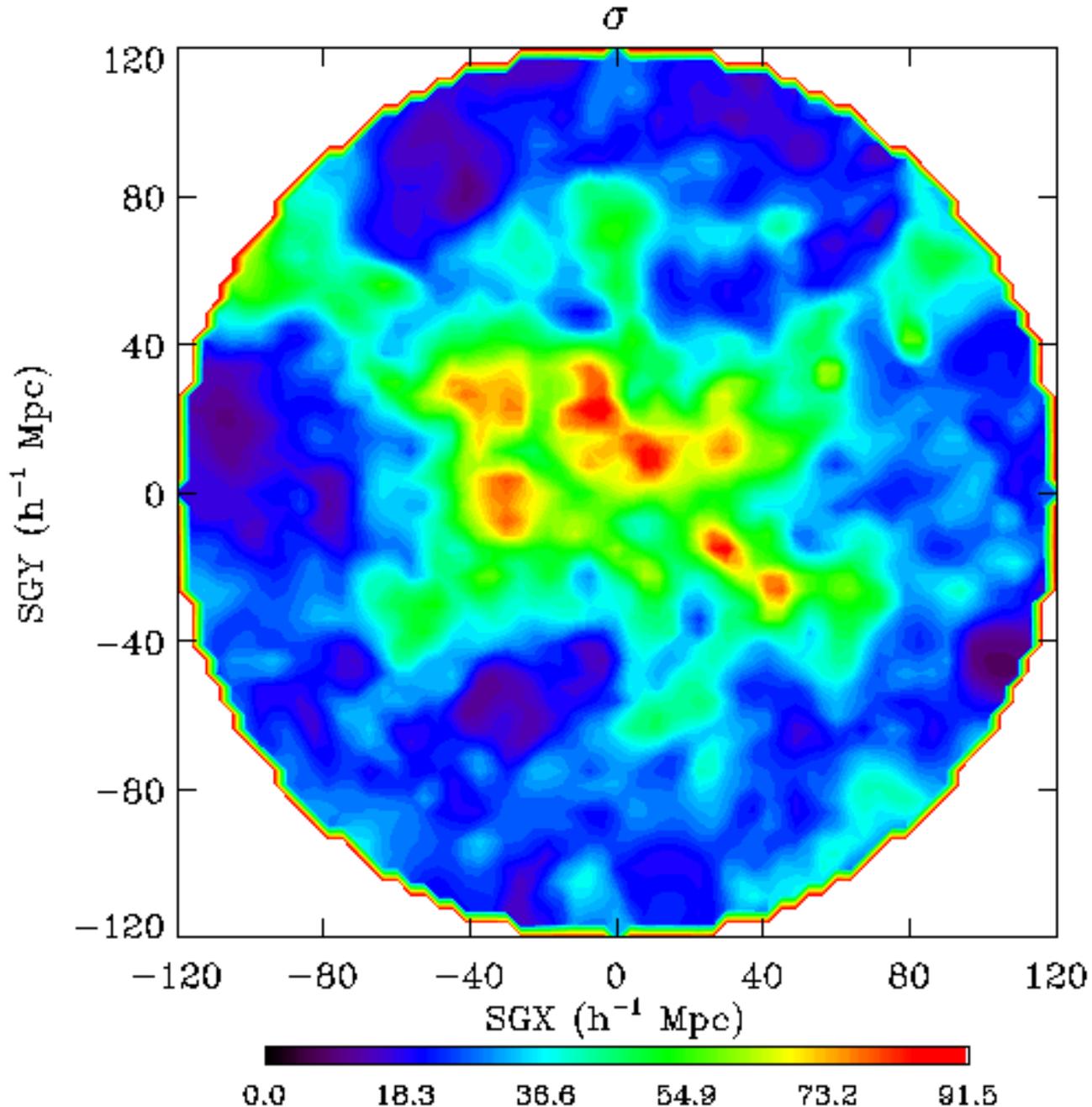}
    \end{minipage}
  \caption{DTFE velocity shear field projected along the
    $z-$supergalactic plane. The slice corresponds to the one
    presented in Fig.~\ref{fig:pscz_den_vel}. We have plotted the
    velocity shear amplitude $\sigma$ in units of the Hubble
    parameter. The colour bar indicates the values of the velocity
    shear in the colour contour map.}
\label{fig:pscz_shear}
\end{figure*} 
\begin{figure*} 
  \begin{minipage}{\textwidth}
    \begin{center}
      \includegraphics[width=\linewidth]{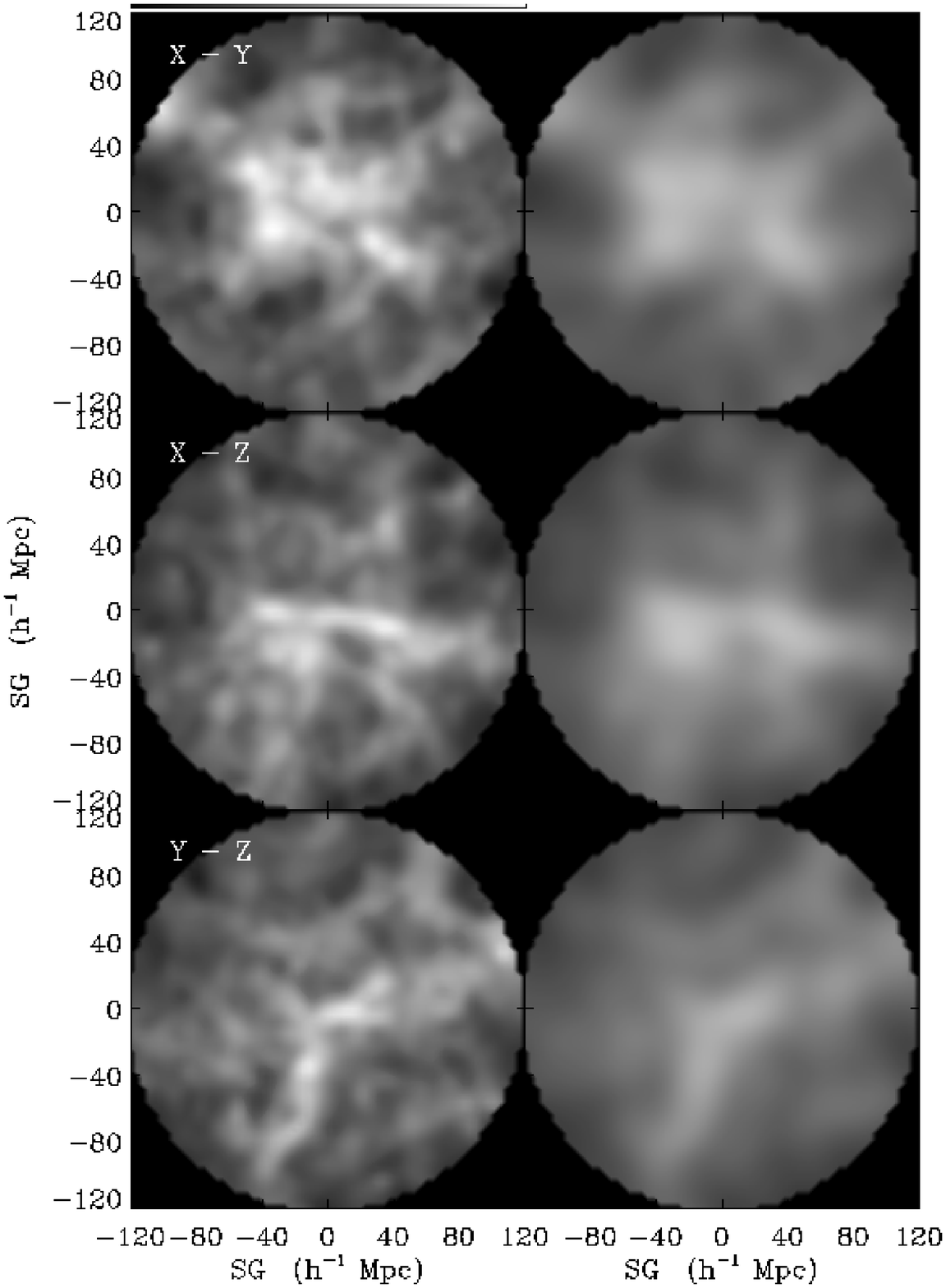}
    \end{center}
  \end{minipage} 
\end{figure*}
\begin{figure*}
  \caption{The DTFE velocity shear field on a smoothing scale of
   $3.8\hmpc$ (lefthand column) and $10 \hmpc$ (righthand column). The
   maps in the consecutive rows concern the three mutually
   perpendicular central supergalactic planes: the X-Y, X-Z and Y-Z
   plane (from top to bottom). Note that the map on a scale of
   $3.8\hmpc$ is uniform only out to a radius of $30\hmpc$, with a
   gradually diminishing resolution towards the outer edge of the
   sample volume. Also compare the images with the density fields in
   fig.~\ref{fig:pscz_den_vel_r3.8_xyz} and
   fig.~\ref{fig:pscz_den_vel_r12_xyz}. The greyscale values are in
   units of the Hubble parameter $H_0$, with their values given in the
   bar at the top of the figure.}
  \label{fig:pscz_shear_3_10_xyz}
\end{figure*} 

\subsection{The Velocity Divergence PDF}
\label{sec:divvpdf}
The righthand panels of figure~\ref{fig:pscz_den-div_pdf} shows the
PDF for the DTFE velocity divergence field. The nice mildly
non-Gaussian character of the pdf (in the top lin-lin plot) argues for
the quality of the DTFE velocity field reconstructions: the DTFE
velocity field interpolation scheme is able to recover to considerable
accuracy the velocity divergence pdf into the nonlinear regime from a
discrete particle distribution.  It may also be noted that the nice
mildly nonlinear character of the DTFE density and velocity divergence
field reconstructions are, at hindsight, an indication for the
validity of the ``linearization'' procedure of \cite{branchini99} into
the early nonlinear regime.

The velocity divergence pdf forms a near perfect mirror image of the
density field pdf (lefthand panel), to be expected on the basis of the
continuity equation through which the velocity divergence and the
density field are related (see next section~\ref{sec:den-divv}). This
relation is reflected in the density and div v pdfs in the lin-lin
plots of the top row and is even more clear when comparing the lin-log
equivalents in the bottom row. DTFE is clearly able to sharply follow
the velocity divergence distribution, both on the negative velocity
divergence side corresponding to infall motions and on the positive
velocity divergence end representative for the outflow from void
region.

It forms a contrast to the rather poor correspondence between the
density and velocity divergence distribution in the corresponding
reconstruction of \cite{branchini99}: the lin-lin plots show that they
do not form each others mirror image. We may also note that its
velocity divergence distribution is considerably broader than that of
the equivalent DTFE pdf. This adheres to a known artefact of gridbased
velocity field interpolations: implicitly these yield mass-weighted
values whose pdfs have wider wings than would be expected for the
distribution of proper volume-weighted quantities in the mildly
nonlinear regime \citep{bernwey96}.

\cite{bernwey96} convincingly demonstrated that the pdf recovered by
Delaunay tessellation (i.e. DTFE) interpolation closely adheres to the
predictions of Eulerian perturbation theory. The velocity divergence
pdf is strongly sensitive to the underlying cosmological parameters,
in particular the cosmological density parameter $\Omega_m$.  By
applying DTFE towards determining the velocity divergence pdf this not
only implies the possibility to get an accurate estimate of $\Omega_m$
but also one which circumvents the usually involved degeneracy between
the cosmic matter density $\Omega_m$ and the bias $b$ between the
matter and galaxy distribution. In a follow-up study \citep{bernwey97}
succesfully tested this on a range of N-body simulations of structure
formation, showing Delaunay interpolation indeed recovered the right
values for $\Omega_m$.

The sensitivity of the pdf to $\Omega_m$ is particularly marked near
the peak of the pdf and in the maximum void expansion rate, ie. the
maximum value of the velocity divergence $\theta_{max}$.  The exact
location of the peak also includes a dependence on cosmological
scenario and level of nonlinearity. To infer $\Omega_m$ extra
assumptions would need to be invoked. It is more
straightforward to read off the value of $\Omega_m$ from the
characteristic sharp cut-off of the pdf on the side of the positive
velocity divergence values. This cut-off relates to the maximum
expansion rate of voids, which is predicted to be
\begin{equation}
\theta_{\rm max}\,=\,1.5\,\Omega^{0.6}\,,
\label{eq:omegapeak}
\end{equation}
\noindent in which we ignore the (rather minor) dynamical influence of
a cosmological constant. The value of 1.5 is the difference in value
of the Hubble parameter in an empty $\Omega=0$ universe and that in an
Einstein-de Sitter Universe $\Omega=1$, reflecting the fact that the
interior of the deepest voids locally mimic the behaviour of an
$\Omega=0$ Universe. Following a related idea \cite{dekelrees94}
obtained a strong lower bound on the value of $\Omega_m \sim 0.3$ on
the basis of an estimate of the value of expansion rate of the nearby
Sculptor void. However, for a proper determination one should seek to
carefully recover the sharp cut-off of the pdf. \cite{bernwey96}
detailed various arguments why a proper determination of the cut-off
is nontrivial. They convincingly demonstrated the success of
tessellation-based methods in outlining the pdf's sharp edge. Here we
infer from the DTFE determined PSCz velocity divergence pdf
(fig.~\ref{fig:pscz_den-div_pdf}) that
\begin{equation}
\Omega_m = 0.35 \pm 0.04\,.
\label{eq:omegacut}
\end{equation}
\noindent This value agrees with the value determined on the basis of
the pdf's peak (eqn.~\ref{eq:omegapeak}.  The $\nabla \cdot {\bf v}$
variation within each individual bins is used to obtain the error
estimate. Given that the input velocity field has a value of
$\beta=0.5$, i.e $\Omega=0.315$ and $b=1$, this forms a confirmation
of the ability of DTFE to self-consistently infer the underlying value
of cosmological parameters, in particular that of $\Omega_m$.

\section{Velocity shear and vorticity}
The map of the amplitude of the shear component in the DTFE velocity
field in the supergalactic X-Y plane is shown in
fig~\ref{fig:pscz_shear}.  The map shows the DTFE estimate of the
amplitude of the shear tensor,
\begin{equation}
{\widehat\sigma}\,\equiv\,\Sigma\, ({\widehat \sigma_{ij}}{\widehat\sigma_{ij}})^{1/2}
\end{equation}
\noindent where ${\widehat \sigma_{ij}}$ is the symmetric traceless
part of the DTFE determined velocity gradient estimate,
\begin{equation}
  \widehat{\sigma_{ij}}\,=\,\frac{1}{2} \biggl\{ \widehat{\frac{\partial v_i}{\partial x_j}} +
  \widehat{\frac{\partial v_j}{\partial x_i}} \biggr\} - \frac{1}{3}
  \big(\widehat{\nabla \cdot {\bf v}} \big) \, \delta_{ij} \,.
  \label{eq:pscz_sh}
\end{equation}
\noindent Shear flows are induced by the intrinsic asphericity of
evolving structures and by the external tidal stresses exerted by the
surrounding (inhomogeneous) large scale matter distribution. These
flows are a manifestation of the tidally induced anisotropic
contraction of matter into planar and filamentary features, locked
into a coherent pattern through highly dense compact clusters which
form at the peaks in the primordial density field.  It forms the basis
for our understanding of the Cosmic Web \citep{bondweb96}, \citep[also
see]{weywhim05}.

The impression provided by the shear field in
fig.~\ref{fig:pscz_shear}is akin to that of the velocity divergence
distribution (fig.~\ref{fig:pscz_div}) and that of the density
distribution (fig~\ref{fig:pscz_den_vel}). Evidently there is a strong
coupling between the density environment and the strength of the shear
field. The velocity shear attains the highest values in and near high
density regions, while the large empty void regions are regions in
which shear flows are hardly relevant with respect to the radial
outflow (ie. the velocity divergence amplitude). This is entirely in
line with theoretical expectations \citep{hoffman86,edbertjain94}.

The spatial shear distribution suggests a very strong tidal force
field in the central region of the Local Universe, in the
supergalactic X-Y plane. Most outstanding are the features in the
supergalactic X-Y plane, in particular the ridge running through the
central Universe from the Pisces Perseus supercluster towards the
Shapley concentration. On the one hand, this impression forms a
telling manifestation of the characteristic quadrupolar pattern of the
local mass distribution, defined by the Hydra-Centaurus supercluster
(the GA region) and the Perseus-Supercluster region.  Even though
undoubtedly real this impression of a strong tidal field in the
central region is partially biased by the effect of a considerably
better spatial resolution in the central $30\hmpc$ of our sample than
in the outer regions and the resulting absence of the (stronger)
higher frequency components present in the center. In order to be able
to assess the impact of this effect we compare the shear field maps
for a smoothing radius of $3.8\hmpc$ with those for $12\hmpc$ in
fig.~\ref{fig:pscz_shear_3_10_xyz}. While it confirms the tendency of
higher shear values in the central region due to the resolved small
scale contributions, the $12\hmpc$ convincingly delineates large scale
patterns responsible for considerable tidal stresses. The low
resolution, large scale, map does indeed demonstrate the reality of
the prominent tidal features in the supergalactic (X-Y)plane. Its
prominent dynamical role is additionally emphasized by the low
resolution X-Z map where the plane stands out as the central edge
marked by high shear values. Interesting features may also be observed
in other regions of the sample volume: an interesting filamentary
extension towards [SGY,SGZ]$\approx[-40,-100] \hmpc$ can be observed
in the Y-Z map.


\subsection{the DTFE Vorticity Map}
Vorticity is not expected to play any role of significance as a result
of the ``linearized'' origin of our PSCz velocities. The top-hat
filter of $R_{TH} = 5\hmpc$ used to defined the velocity sample is an
ensurance for the linearity of practically all sample velocity data.

The presence of vorticity in our DTFE reconstructions should therefore
provide us with a reasonably good impression of the influence of
systematic artefacts in our maps. Regions populated with considerable
large or abundant clusters (e.g. Great Attractor, Pisces-Perseus), do
retain some measure of vorticity, even after smoothing. Also the DTFE
method itself also introduces a small spurious vorticity component as
a result of its linear interpolation scheme. A careful analysis of
this artefact in our maps, and the effects for the DTFE velocity
reconstruction, will be presented in \cite{romdiaz07}.

\section{Summary and Discussion}
The Delaunay Tessellation Field Estimator \citep[DTFE,
see][]{schaapwey2000} has been applied to a combined analysis of the
density and velocity flow field in the local Universe. The prime
objective of this study has been the production of optimal resolution
three-dimensional maps of the volume-weighted velocity and density
fields throughout the nearby Universe, the basis for a detailed study
of the structure and dynamics of the cosmic web at each level probed
by underlying galaxy sample. In order to have a reasonably complete
sample of galaxy peculiar velocities throughout the surrounding local
Universe we used the \pscz galaxy redshift catalog translated into
galaxy positions and velocities by means of the method I linearization
process of \cite{branchini99}.

The spatial galaxy distribution of the \pscz catalog defines a Voronoi
and Delaunay tessellation. These epitomize the most pure locally
defined division of space in the \pscz volume. The self-adaptive
nature of the Voronoi/Delaunay tessellations concerns both spatial
resolution and local geometry. The high level sensitivity to the local
point distribution is exploited by DTFE \citep{schaapwey2000} to
produce an estimate of the local density at each sample point. The
Delaunay tessellation is subsequently used as adaptive
multidimensional spatial interpolation grid.  The construction of the
Voronoi and Delaunay tessellation is the most demanding task of the
DTFE routine. Once this has been accomplished, they can be used for
both the density and velocity field (and basically any relevant
field).

Because DTFE is based upon linear interpolation and involves constant
field gradients within each Delaunay tetrahedron, the analysis of the
cosmic velocity field automatically yields maps of the velocity
divergence, shear and even vorticity. The spatial distribution of
these quantities in DTFE is marked by a discontinuous zeroth order map
marked by Delaunay regions in which divergence, shear or vorticity has
a constant value. To sensibly relate the DTFE velocity divergence and
shear fields it is necessary to smooth over a scale comparable to that
of the corresponding Delaunay cells.

Earlier studies \citep{schaapwey06a} have made clear that DTFE is
particularly optimalized for the analysis of mass distributions marked
by one or more of the essential aspects of the nonlinear cosmic web of
the Megaparsec Universe \citep{bondweb96}.  One of the main advantages
of DTFE is its large dynamic range, allowing it to trace small scale
structures along with the large scale environment in which they are
embedded. The spatial and morphological adaptivity of DTFE is ideally
suited for dissecting the essential components of the nonlinear
weblike cosmic matter distribution. It resolves a hierarchically
structured matter distribution to the smallest possible resolution
scale set by the particle number density. Perhaps most outstanding is
the ability of DTFE to retain the morphology cq. shape of the features
and patterns in the matter distribution: the characteristic
anisotropic filamentary and planar features of the {\it Cosmic web}
are fully reproduced in the continuous DTFE density field
\citep{schaapwey2000}. Also the third major characteristic of the
nonlinear Megaparsec universe, the dominant existence of near-empty
voids, is resolved by DTFE. Both their flat internal density
distribution as well as their sharp outline and boundary are recovered
in detail, while shot noise in these sparsely sampled regions tends to
be suppressed.

The maps of the \pscz local Universe reveal a sharply defined density
field in which many familiar structures and features can be recognized
at an optimal spatial resolution.  The region around the Local Group
and its filamentary extension towards the Hydra-Centaurus and Shapley
complexes as well as the Pisces Perseus supercluster are prominently
resolved in the DTFE density fields. Moreover, these features are as
clearly and sharply defined in the corresponding velocity
flows. Perhaps the most outstanding features in the velocity vector
fields are the radially expanding void regions.  In comparison to
earlier published maps of the density field and velocity flow in the
\pscz sample volume \citep{branchini99,schmoldt1999} the DTFE maps
have a considerably sharper defined appearance. While conventional
maps suppressed small scale details and were at loss in undersampled
void regions, in particular wrt the interpolation of the velocity
field, the DTFE fields reveal a beautifully textured flow field.  The
DTFE flows are marked by prominent bulk flows, shear flows and radial
inflow (Coma cluster). Arguably the most outstanding and unique
feature of the DTFE maps is the sharply defined radial outflow regions
in and around underdense voids, marking the dynamical importance of
voids in the Local Universe.

The present study includes two specific aspects and challenges for
DTFE: (1) the application of DTFE to reconstruct the density field as
well as the corresponding velocity field in the same cosmic volume and
(2) the gradually diminishing spatial resolution as a function of
radial distance as a consequence of the flux-limited nature of the
\pscz sample.

While one may correct estimated density values by taking into account
the well-defined radial selection function, it is not possible to
correct for the loss in spatial resolution: the lower galaxy sampling
density goes along with a loss, at increasing distances, in spectral
coverage of the maps. A full recovery would have to involve
pre-conceived notions on the cosmological structure formation
scenario, cq. the power spectrum. The relatively prominent central
values in the density, velocity divergence and shear maps are a
reflection of this effect. On the other hand, the velocity field
itself -- less affected by the lack of high frequency spectral power
at large distances -- for all practical purposes appears to be
uniformly covered by the DTFE reconstruction.  The sharply outlined
edges of superclusters and the radial outflow of voids are found
throughout the \pscz volume.

We have addressed the relationship between the density field and
velocity field by correlating the density values with the local
velocity divergence and shear. Overall, the linear density-velocity
divergence relationship is accurately reproduced. This reflects the
origin of the galaxy velocities, having been computed from a redshift
map through a linearization procedure. These simple quantitative
relations do not express the spatial coherence and correlations within
the velocity field. The tight spatial correspondence between density
features and velocity flows in the DTFE maps -- voids with radial
outflow, elongated and flattened superclusters with bulk and shear
flows -- does demonstrate the remarkable adaptive nature of the DTFE
technique and its promise of understanding the dynamics of the cosmic
web. The maximum expansion rate of voids defines a sharp cutoff in the
velocity divergence pdf. It enabled us to test the self-consistency of
the DTFE method. Indeed it confirmed the value of $\Omega_m \approx
0.35$ of the cosmology underlying the galaxy velocity data.

For furthering our insight into the velocity field in the nearby
universe it would be desirable to be able to probe quasi-linear
contributions to the velocity field. Techniques like the FAM$-z$ of
\cite{branchini2002}, applied in combination with the DTFE algorithm,
would offer a potentially promising approach. While this study has
shown the potential of DTFE towards a succesfull analysis of velocity
fields it concerns a test on the basis of a sample of pre-processed
galaxy velocities. This includes the assumption that the velocity
field linearization \citep{branchini99} has succesfully dealt with
redshift-space distortions and sampling effects.

Ultimately we will seek to develop a DTFE-based formalism to directly
analyze samples of galaxy peculiar velocities on the basis of the
``raw'' observational data. The challenge is found in a few
aspects. One aspect concerns errors and sampling effects.  A second
aspect involves redshift distortions. A third complicating aspect
involves fundamental assumptions behind the velocity field
reconstruction procedure, such as the requirement of laminarity of the
flow. An extensive study of various sampling effects, such as
decreased sampling density and inhomogeneous sampling, on
simulation-based data do show that the noise level of the DTFE
velocity reconstructions does increase but that this does not involve
any systematic shifts \citep{emiliophd}. Given the acclaimed optimal
triangulation properties of Delaunay grids underlying the development
of DTFE this may not come as a surprise. The same study involved tests
of the effects of a variety of different error contributions.  They
induced similar effects as a less balanced galaxy sampling. These
results will be presented in an extensive study and report on error
and sampling effects in a forthcoming paper. More challenging issues
concern the physically more profound aspects of redshift distortions
and higher-order flow characteristics. As a first-order version of the
wider class of natural-neighbour interpolation schemes
\citep{sibson1980, sibson1981,watson1992}, DTFE may be extended to a
higher-order natural neighbour interpolation scheme based upon the
Delaunay tessellation of a galaxy sample. Such a formalism would
enable a more complex treatment of nonlinear velocity flows, possibly
even of multistream flows. Already we have been working towards the
implementation of these higher-order interpolation schemes
\citep[see][for a review on the basics]{weyschaap2007}, and we plan to
look into the possibility on the basis of routines made available
through \cgal \footnote{\cgal is a \texttt{C++} library of algorithms
and data structures for Computational Geometry, see
\url{www.cgal.org}.}. On the other hand, highly mixed virialized
regions will remain beyond the scope of these interpolation
techniques. This may hardly be considered a constraint given the fact
that for the foreseeable future the focus of nearly all galaxy
peculiar velocity surveys will be the dynamics of large scale
structures.  The inversion from redshift to real space data may be
facilitated and enabled by using the virtues of a Delaunay grid for
the solution of PDEs. The use of Delaunay tessellations as grids for
the numerical solution of PDEs has been first described by
\cite{braunsambridge1995}, their success provides substantial
imperative for the formulation of the redshift space to real space
inversion following these lines.

In the meantime DTFE has been elaborated upon through the development
of a number of specific feature detecting techniques. The MMF
\citep{aragonmmf2007} forms a highly promising tool for selection and
identification of filamentary and planar structures and their spatial
relationship with neighbouring structures. The watershed based void
detection algorithm developed by \cite{platen2007} has been applied
towards the analysis of the hierarchical evolution of voids
\citep[see][]{shethwey2004}.  In all cases, the fully adaptive and
optimal tessellation based characteristics of DTFE form the crucial
starting point.


\section*{Acknowledgments}
We are very grateful to E. Branchini for providing us with the PSCz
sample of galaxy positions and linearized velocities. Also we wish to
thank W. Schaap for the use of DTFE routines and for useful comments
and F. Bernardeau for crucial advice. We greatly benefited from
discussions about velocity fields and Delaunay tessellations with
Bernard Jones and Manolis Plionis.  Finally, we wish to thank the
referee, E. Branchini, for useful suggestions and comments. ERD has
been supported by the Golda Meier Fellowship at the Hebrew University,
and HST/AR 10976, NASA/LTSA 5-13063, NSF/AST 02-06251, and NASA/ATP
NNG06-GJ35G. Part of this work was finished while enjoying the
hospitality of the NOA in Athens, which we gratefully acknowledge.


\bibliographystyle{mn2e}

\label{lastpage}

\end{document}